\documentclass{jfm}

\usepackage{cite}
\usepackage{url}

\usepackage{graphicx}
\usepackage{natbib}
\usepackage{amssymb,amsmath}

\usepackage{graphicx, amsmath, esint, bm}
\usepackage{epstopdf, epsfig}

\newcommand{\DETAILS}[1]{}

\newcommand{\I}{\mathrm{i}}
\newcommand\D{\mbox{d}}

\newcommand{\R}{\mathbb R}
\newcommand{\C}{\mathbb C}
\newcommand{\e}{\eqref}

\newcommand \bfr{{\bf r}}

\ifCUPmtlplainloaded \else
  \checkfont{eurm10}
  \iffontfound
    \IfFileExists{upmath.sty}
      {\typeout{Found AMS Euler Roman fonts on the system,
                   using the 'upmath' package.}%
       \usepackage{upmath}}
      {\typeout{Found AMS Euler Roman fonts on the system, but you
                   dont seem to have the}%
       \typeout{'upmath' package installed. JFM.cls can take advantage
                 of these fonts, if you use 'upmath' package.}%
       \providecommand\upi{\upi}%
      }
  \else
    \providecommand\upi{\upi}%
  \fi
\fi

\ifCUPmtlplainloaded \else
  \checkfont{msam10}
  \iffontfound
    \IfFileExists{amssymb.sty}
      {\typeout{Found AMS Symbol fonts on the system, using the
                'amssymb' package.}%
       \usepackage{amssymb}%
       \let\le=\leqslant  
         
      }{}
  \fi
\fi

\ifCUPmtlplainloaded \else
  \IfFileExists{amsbsy.sty}
    {\typeout{Found the 'amsbsy' package on the system, using it.}%
     \usepackage{amsbsy}}
    {\providecommand\boldsymbol[1]{\mbox{\boldmath $##1$}}}
\fi

\shorttitle{Dynamics of Poles: New Constants of Motion.}
\shortauthor{A. I. Dyachenko, S. A. Dyachenko, P. M. Lushnikov, and
V. E. Zakharov}

\title{Dynamics of Poles in $2$D Hydrodynamics with Free Surface: New Constants of Motion}

\author{ A. I. Dyachenko\aff{1,2},
   S. A. Dyachenko\aff{3,4},
         P. M. Lushnikov\aff{1,5}\corresp{\email{plushnik@math.unm.edu}}, \and V. E. Zakharov\aff{1,2,6}
 }

\affiliation{ \aff{1}Landau Institute for Theoretical Physics,
Chernogolovka, 142432, Russia  \aff{2}Center for Advanced Studies,
Skoltech, Moscow, 143026, Russia \aff{3}Department of Mathematics,
University of Illinois at Urbana-Champaign, Urbana, IL 61801, USA
\aff{4}Department of Applied Mathematics, University of Washington,
Seattle, WA 98195, USA \aff{5}Department of Mathematics and
Statistics, University of New Mexico, Albuquerque, NM 87131, USA
\aff{6}Department of Mathematics, University of Arizona, Tucson, AZ
85721, USA }

\begin{document}

\maketitle

\centerline{Dated: September 24, 2018}

\begin{abstract}
We address a problem of potential motion   of ideal incompressible
fluid with a free surface and infinite depth in two dimensional
geometry. We admit a presence of gravity forces and surface tension.
A time-dependent conformal mapping $z(w,t)$ of the lower complex
half-plane of the variable $w$ into the area filled with fluid is
performed with the real line of $w$ mapped into the free fluid's
surface. We study the dynamics of singularities of   both $z(w,t)$
and the complex fluid potential $\Pi(w,t)$ in the upper complex
half-plane of $w$. We show  the existence of solutions with an
arbitrary finite number $N$ of  complex poles in   $z_w(w,t) $ and
$\Pi_w(w,t)$ which are the derivatives of   $z(w,t) $ and $\Pi(
w,t)$ over $w$. We stress that these solutions are not purely
rational because they generally have branch points at other
positions of the upper complex half-plane.    The orders of poles
can be arbitrary for zero surface tension while  all orders are even
for nonzero surface tension. We find that the residues of
$z_w(w,t)$  at these $N$ points are new, previously unknown
constants of motion, see also Ref. V. E. Zakharov and A. I.
Dyachenko, arXiv:1206.2046 (2012) for the preliminary results.  All
these constants of motion commute with each other in the sense of
underlying Hamiltonian dynamics. In absence of both gravity and
surface tension, the residues of      $\Pi_w(w,t)$ are also the
constants
 of motion while nonzero gravity $g$ ensures
  a trivial linear dependence of  these residues on time. A Laurent series expansion of both    $z_w(w,t) $ and $\Pi_w(w,t)$  at each poles position reveals an existence of
  additional integrals of motion for  poles of
 the second  order.   If all poles are simple then the number of independent real integrals of motion is $4N$ for  zero gravity and $4N-1$ for nonzero gravity. For the second order poles we found $6N$ motion integral for zero gravity and $6N-1$ for nonzero gravity. We suggest that the existence of these nontrivial constants of motion provides an argument in support of the conjecture of complete integrability of free surface hydrodynamics in deep water. Analytical results are solidly supported by high precision numerics. \end{abstract}


\section{Introduction and basic equations }
 \begin{figure}
\includegraphics[width=0.859\textwidth]{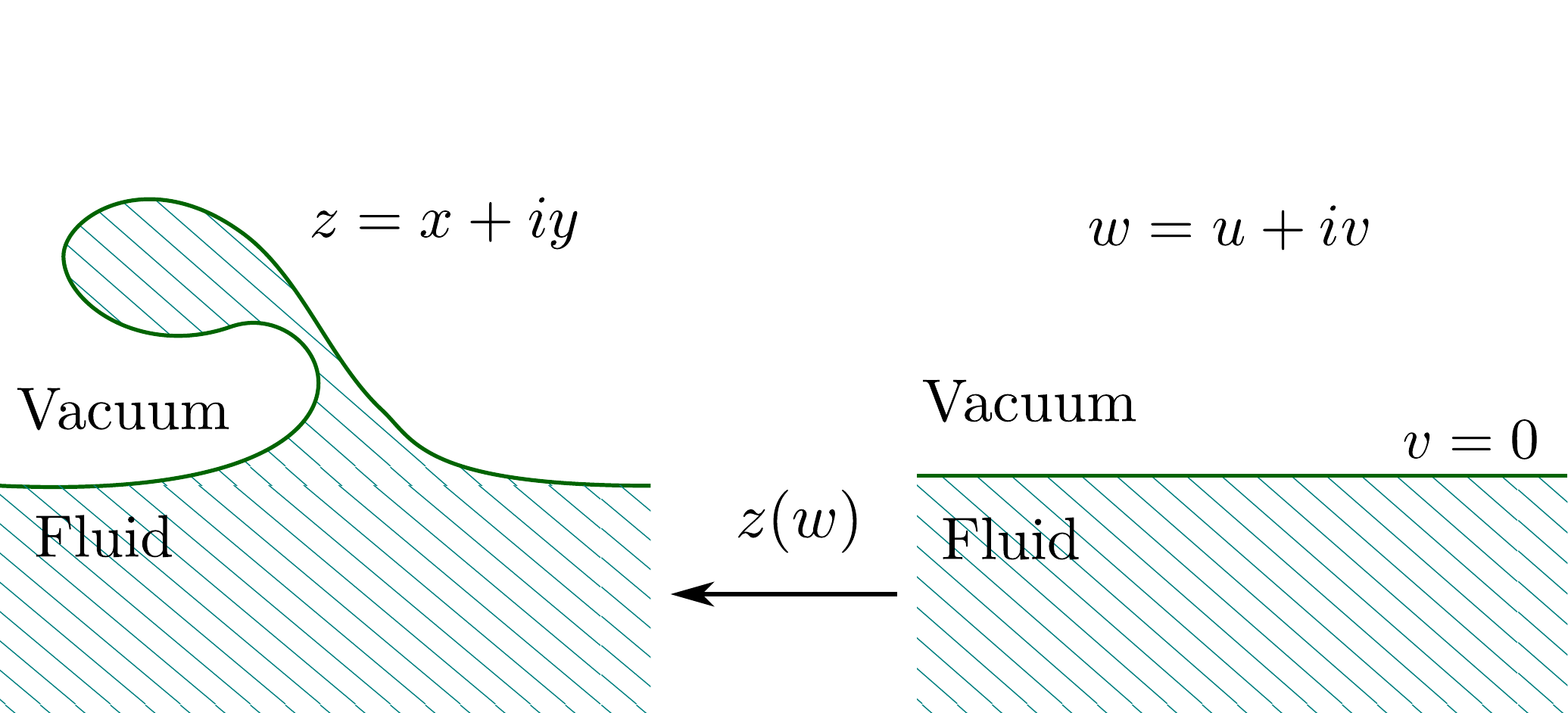}
\caption{ Shaded area represents the domain occupied by fluid in the
physical plane $z=x+\I y $ (left) and the same domain in   $w=u+\I
v$ plane  (right). Thick solid lines correspond to the fluid's free
surface.} \label{fig:schematic1}
\end{figure}

We consider two-dimensional potential motion of
ideal incompressible fluid with free surface of infinite depth.
Fluid occupies the infinite region $-\infty < x < \infty$ in the
horizontal direction $x$ and extends down to $y\to -\infty$ in the
vertical direction $y$   as schematically shown on the left panel of
Fig. \ref{fig:schematic1}. We assume that fluid is unperturbed both at
$x\to \pm \infty  $ and $y\to -\infty$.

We use a time-dependent conformal mapping
\begin{equation} \label{zwdef}
z(w,t)=x(w,t)+\I y(w,t)
\end{equation}
of the lower complex half-plane $\mathbb{C}^-$ of the auxiliary complex variable
%
$w\equiv u+\I v, \quad -\infty<u<\infty,
$
%
into the area in $(x,y)$ plane occupied by the fluid. Here the real line $v=0$ is mapped into the fluid free surface (see Fig. \ref{fig:schematic1}) and $\mathbb{C}^-$ is defined by the condition  $-\infty<v\le0$.
 Then the time-dependent fluid free surface is represented in the parametric form as
\begin{equation} \label{xyu}
x=x(u,t), \ y=y(u,t).
\end{equation}
A decay of perturbation of fluid beyond flat surface at  $x(u,t)\to \pm \infty$  and/or $y\to -\infty$ requires that %
\begin{equation} \label{zlimit}
z(w,t)\to w \ \text{for} \ |w|\to\infty, \ w\in\C^-.
\end{equation}

The conformal mapping \e{zwdef} imply that $z(w,t)$ is the analytic function of   $w\in\mathbb{C^-} $ and  %
\begin{equation} \label{zwconformal}
z_w\ne 0 \ \text{for any} \  w\in\mathbb{C^-}.
\end{equation}

Potential  motion means
 that a velocity ${\bf v}$    of fluid is determined by a
velocity potential $\Phi(\bfr,t)$ as ${\bf v}= \nabla \Phi$ with $\nabla\equiv(\frac{\p}{\p x},\frac{\p}{\p y})$.  The
incompressibility condition $\nabla \cdot {\bf v} = 0$ implies the
Laplace equation
\begin{align} \label{laplace}
\nabla^2 \Phi = 0
\end{align}
inside fluid, i.e. $\Phi$ is the harmonic function inside fluid.
 Eq. \e{laplace} is supplemented with  a decaying boundary condition (BC) at infinity,  %
\begin{equation} \label{phiinfinity}
\nabla\Phi\to 0 \ \text{for } \ |x|\to \infty \ \text{or} \ y\to - \infty.
\end{equation}
 The harmonic conjugate of $\Phi$ is a  stream  function $\Theta$ defined by%
\begin{equation} \label{Thetadef}
\Theta_x=-\Phi_y \ \text{and} \ \Theta_y=\Phi_x.
\end{equation}
Similar to Eq. \e{phiinfinity}, we set without loss of generality a zero  Dirichlet BC for $\Theta$ as

\begin{equation} \label{Dirichlet2}
\Theta\to 0 \  \text{} \ \text{for } \ |x|\to \infty \ \text{or} \ y\to - \infty.
\end{equation}

 We define a complex velocity potential $\Pi(z,t)$~as%
\begin{equation} \label{ComplexPotentialdef}
\Pi=\Phi+\I \Theta.
\end{equation}
%
%
is the complex coordinate. Then Eqs. \e{Thetadef} turn into
Cauchy-Riemann equations ensuring the analyticity of $\Pi(z,t)$ in the domain of $z$ plane occupied by the fluid. A physical velocity with the components
$v_x$ and $v_y$ (in $x$ and $y$ directions, respectively) is obtained from $\Pi$ as $\frac{d\Pi}{dz}=v_x-\I v_y$.
The conformal mapping \e{zwdef} ensures that the function $\Pi(z,t)$~ %
\e{ComplexPotentialdef} transforms into $\Pi(w,t)$ which is analytic function of $w$ for $w\in\mathbb{C^-}$ (in the bulk of
fluid).
Here and below we abuse the notation and use the  same symbols for functions of either  $w$ or $z$     (in other words, we  assume that e.g. $\tilde \Pi(w,t)= \Pi(z(w,t),t) $ and remove $\tilde ~$ sign).
The conformal transformation    \e{zwdef} also ensures   Cauchy-Riemann equations $
\Theta_u=-\Phi_v, \quad \Theta_v=\Phi_u $
  in $w$ plane.

BCs at the free surface  are time-dependent and consist of kinematic and dynamic BCs.
A kinematic BC ensures that  free surface moves with the normal velocity component $v_n$ of fluid particles at the free surface.   Motion of the free surface is determined by a time derivative of the parameterization  \e{xyu}  while the kinematic BC is given by a projection into the normal direction as
\begin{equation} \label{kinematicu0}
{\bf n}\cdot\left(x_t,y_t \right )=v_n\equiv{\bf n}\cdot\nabla \Phi|_{x=x(u,t),\ y=y(u,t)},
\end{equation}
where %
${\bf n}=\frac{(-y_u,x_u)}{(x_u^2+y_u^2)^{1/2}}
$
is the outward unit normal vector to the free surface and subscripts here and below means partial derivatives, $x_t\equiv\frac{\p x(u,t)}{\p t}$ etc.

Eq. \e{kinematicu0}  results in a compact expression  %
\begin{equation}\label{fullconformal1}
 y_tx_u  -x_t  y_u =- \hat H \psi_u
\end{equation}
for the kinematic BC as was found in Ref. \cite{DKSZ1996}, see also Ref. \cite{DyachenkoLushnikovZakharovJFM2019} for more details. Here    %
\begin{equation} \label{dirichletpsi}
\psi(u,t)\equiv\Phi(u,v,t)|_{v=0}
\end{equation}
is the Dirichlet BC for $\Phi$ at the free surface and
\begin{equation} \label{Hilbertdef}
\hat H f(u)=\frac{1}{\upi} \text{p.v.}
\int^{+\infty}_{-\infty}\frac{f(u')}{u'-u}\D u'
\end{equation}
is the Hilbert
transform with $\text{p.v.}$ meaning a Cauchy principal value of the integral.
Real and imaginary parts of both $z$ and $\Pi$ at $v=0$ are related through $\hat H$ as follows
\begin{equation} \label{xHy}
\tilde    x\equiv x-u=-\hat Hy, \quad \hat H x=y
\end{equation}
and
\begin{equation} \label{PhiTheta1}
\Theta |_{w=u}= \hat H \psi, \quad \psi =- \hat H \Theta|_{w=u},
\end{equation}
see e.g. Appendix A of Ref.
\cite{DyachenkoLushnikovZakharovJFM2019}. Thus it is sufficient to
find $y(u,t)$ and $\psi(u,t)$ while $x(u,t)$ and $\Theta(u,t)$ can
be recovered from Eqs. \e{xHy} and \e{PhiTheta1}.

 A dynamic BC is given by the time-dependent Bernoulli equation (see e.g. \citet{LandauLifshitzHydrodynamics1989})
at the free surface,%
\begin{align} \label{dynamic1}
\left.\left(\Phi_t +
 \dfrac{1}{2}\left(\nabla \Phi\right)^2+gy\right)\right|_{x=x(u,t),\ y=y(u,t)}  = -P_\alpha,
\end{align}
 where $g$ is the acceleration due to gravity and %
$P_\alpha=-\frac{\alpha(x_uy_{uu}-x_{uu}y_u) }{(x_u^2+y_u)^{3/2}}
$ 
  is the pressure jump  at the free
surface due to the surface tension coefficient $\alpha$. Here without loss of generality we assumed that pressure is zero above the free surface (i.e. in vacuum).
All results below apply both to surface gravity wave case ($g>0$) and Rayleigh-Taylor problem $(g<0)$. We also consider a particular case $g=0$ when inertia forces  well exceed
 gravity force.

Eq. \e{dynamic1} can be transformed into
\begin{equation}\label{fullconformal2}
\psi_t y_u - \psi_u y_t + gyy_u =- \hat H \left (\psi_t x_u -
\psi_ux_t + gyx_u \right )-\alpha\frac{\p  }{\p u}\frac{x_u}{|z_u|}+\alpha\hat H \frac{\p  }{\p u}\frac{y_u}{|z_u|},
\end{equation}
thus representing the dynamic BC in the conformal variables, see
Refs. \cite{DyachenkoLushnikovZakharovJFM2019} for  details of such
transformation.

Eqs.
 \e{fullconformal1},\e{xHy} and \e{fullconformal2}  form a closed set of equations which is equivalent to Euler equations for dynamics of ideal fluid with free surface. The idea of using time-dependent conformal transformation like
\e{zwdef} to address systems equivalent/similar to  Eqs.
 \e{fullconformal1},\e{xHy} and \e{fullconformal2}   was
exploited by several authors
including~\cite{Ovsyannikov1973,MeisonOrzagIzraelyJCompPhys1981,TanveerProcRoySoc1991,TanveerProcRoySoc1993,DKSZ1996,ChalikovSheininAdvFluidMech1998,ChalikovSheininJCompPhys2005,ChalikovBook2016,ZakharovDyachenkoVasilievEuropJMechB2002}.
We follow the analysis of Refs.
\cite{ZD2012,DyachenkoLushnikovZakharovJFM2019}
  which found that  Eqs. \e{fullconformal1},\e{xHy} and \e{fullconformal2} can be explicitly solved for the time derivatives $y_t, \psi_t$ and rewritten  in the non-canonical Hamiltonian form
\begin{equation} \label{implectic2}
 {\bf Q}_t=\hat R\frac{\delta H}{\delta {\bf Q}}, \quad {\bf Q}\equiv\begin{pmatrix}y \\
\psi
\end{pmatrix}
\end{equation}
for the Hamiltonian variables  $y(u,t)$ and $\psi(u,t),$   where%
$\hat R=\hat \Omega^{-1}=\begin{pmatrix}0 & \hat R_{12} \\
\hat R_{21} &  \hat R_{22} 
\end{pmatrix}
$
is $2\times 2$ skew-symmetric matrix operator with the components
\begin{equation}\label{Rmatrdef}
\begin{split}
& \hat R_{11}q=0,\ \hat R_{12}q=\frac{x_u}{J}q-y_u\hat H\left ( \frac{q}{J}\right ), \\
& \hat R_{21}q=-\frac{x_u}{J}q-\frac{1}{J}\hat H\left (y_u q\right ), \quad  \hat R_{21}^\dagger=-\hat R_{12},\\
 & \hat R_{22}q=-\psi_u\hat H\left ( \frac{q}{J}\right ) -\frac{1}{J}\hat H\left (\psi_u q\right ),\quad  \hat R_{11}^\dagger=-\hat R_{11}.
\end{split}
\end{equation}
We call $\hat R=\hat \Omega^{-1}$ by the ``implectic"  operator (sometimes such type of inverse of the
symplectic operator  is also called by the co-symplectic operator, see e.g. Ref. \cite{WeinsteinJDiffGeom1983}).
Here the Hamiltonian $H$ is the total energy of fluid (kinetic plus potential energy in the gravitational field and surface tension energy) which is written in terms of the Hamiltonian variables as
\begin{equation} \label{Hpsiyonly}
H=-\frac{1}{2}\int\limits^\infty_{-\infty}\psi\hat H\psi_u\D u+\frac{g}{2}\int \limits_{-\infty}^{\infty} y^2\,(1-\hat H y_u)\D  u+\alpha\int\limits^{\infty}_{-\infty}\left (\sqrt{(1-\hat H y_u)^2+y_u^2}-1+\hat H y_u\right )\D u.
\end{equation}

Eqs. \e{implectic2} allows to define the Poisson bracket (see Ref.
\cite{DyachenkoLushnikovZakharovJFM2019})
\begin{equation}\label{PoissonBracketsRDef}
\begin{split}
& \{F,G\}
=\int \limits^{\infty}_{-\infty}\D u\left ( \frac{\delta F}{\delta y }\hat R_{12}  \frac{\delta G}{\delta \psi } +\frac{\delta F}{\delta\psi }\hat R_{21}  \frac{\delta G}{\delta y }+\frac{\delta F}{\delta \psi }\hat R_{22}  \frac{\delta G}{\delta \psi }\right )
\end{split}
\end{equation}
which allows to rewrite Eq. \e{implectic2} in terms of  Poisson mechanics as %
\begin{equation}\label{hamiltoniancanonicalpoisson}
\begin{split}
 {\bf Q}_t=\{{\bf  Q},H\}.
 \end{split}
\end{equation}
 Thus a functional $F$ is the constant of motion of Eq. \e{hamiltoniancanonicalpoisson}
provided $\{{ F},H\}=0.$

The Hamiltonian system \e{implectic2}-\e{hamiltoniancanonicalpoisson} is the generalization of the results of  Ref. \cite{Zakharov1968}. It was conjectured in Ref. \cite{DZ1994}  that the system
\e{xHy}, \e{fullconformal1} and \e{fullconformal2}  is completely integrable at least for the case of
the zero surface tension. Since then the arguments {\it pro} and
{\it contra} were presented, see e.g. Ref. \cite{DyachenkoKachulinZakharovJETP2013eng}. Thus this
question is still open.

The system
 \e{fullconformal1},\e{xHy} and \e{fullconformal2}   has an infinite number of degrees of freedom. The most important feature of integrable systems
   is the existence of ``additional"
constants of motion which are different from ``natural" motion constants (integrals) (see Refs. \cite{ArnoldClassicalMechanics1989,ZakharovFaddeevFunktAnalPril1971eng,NovikovManakovPitaevskiiZakharovbook1984}). For the system
 \e{fullconformal1} , \e{xHy} and \e{fullconformal2},   the natural integrals are the energy $H$  \e{Hpsiyonly}, the total mass of fluid, %
and the horizontal component of the momentum. For $g=0,$  the
vertical component of momentum is  also the integral of motion. See
Ref. \cite{DyachenkoLushnikovZakharovJFM2019} for the explicit
expressions for these natural integrals.

In this paper we show that  the system
 \e{fullconformal1},\e{xHy} and \e{fullconformal2} has   a number of additional constants of motion.
 We cannot so far determine/estimate a total number of these constants. Instead we show examples of initial data such that
 the system has almost obvious, very simply constructed additional constants. We must stress that  the number of known additional constants depends
 so far on the choice of initial data and can be made arbitrary large for the specific choices of initial data.
Some of these  new integrals of motion are  functionals  $y$ only.
It  follows from Eq. \e{PoissonBracketsRDef} that any functionals
$F$ and $G,$ which  depend only on $y$,  commute with each other,
i.e.  $\{{ F},G\}=0.$ We suggest that the existence of such
commuting integrals of motion might be a sign of the Hamiltonian
integrability of the free surface hydrodynamics. Such conjecture is
in agreement with the history of the discovery of the Hamiltonian
integrability of Korteweg–de Vries equation, nonlinear Schr\"odinger
and many other partial differential equations, see Refs.
\cite{GardnerGreeneKruskalMiuraPRL1967,ZakharovShabatJETP1972,ArnoldClassicalMechanics1989,ZakharovFaddeevFunktAnalPril1971eng,NovikovManakovPitaevskiiZakharovbook1984}).


 Plan of the paper is the following.
 In Section \ref{DynamicalequationsComplex} we introduce dynamic equations in the complex form for  another
  unknowns $R$ and $V$ and consider an analytical continuation of solution into the upper complex half plane.
   Section \ref{sec:poles} discusses non-persistence of pole solutions in  both $R$ and $V$ variables within arbitrary small time while addressing that power law branch points are persistent.
    New constants of motion for gravity case but with zero
surface tension are found in Section \ref{sec:Newconstantsmotion}
for solutions of full hydrodynamic equations with simple complex poles in the original
variables $z_w$ and $\Pi_w$. Section
\ref{sec:Anotherveiwmotionconstants} provides another view of the
new motion constants. Section
\ref{sec:Newconstantsmotioncapillarity} identifies  new
constants of motion to   nonzero surface tension and second order
poles.  Section \ref{sec:Kelvintheorem} discusses a global analysis
for analytical continuation into multi-sheet Riemann surfaces and
introduce a Kelvin theorem for phantom hydrodynamics. Section \ref{sec:numericalsimulations} provides a  brief description of our numerical methods for simulation of free surface dynamics by spectrally accurate adaptive mesh refinement approach  and a procedure for recovering of the structure of the complex singularities above fluid's surface.   Section
\ref{sec:recoveringsingularities} is devoted to the numerical
results on free surface hydrodynamics  simulations which provides a detailed
verification  of results of all other sections. Section
\ref{sec:conclusion} gives a summary of obtained results and
discussion of future directions.

\section{Dynamic equations in the complex form and analytical continuation of solution into the upper complex half plane} \label{DynamicalequationsComplex}

Dynamical Eqs.
 \e{fullconformal1},\e{xHy} and \e{fullconformal2}  are defined on the real line $w=u$ with the analyticity of $z(w,t)$ and $\Pi(w,t)$ in $w\in \C^-$  taken into account through the Hilbert operator $\hat H.$ In this paper we consider also analytical continuation of these functions into the upper complex half plane $w\in\C^+.$    Both $z(w,t)$ and $\Pi(w,t)$ has time dependent complex singularities for $w\in \C^+$.

Using the Hilbert operator $\hat H$ \e{Hilbertdef}, we introduce the  operators
\begin{equation} \label{Projectordef}
\hat P^-=\frac{1}{2}(1+\I \hat H)  \quad\text{and}\quad  \hat P^+=\frac{1}{2}(1-\I \hat H)
\end{equation}
which are the projector operators of a  function $q(u)$ defined at the real
line $w=u$ into  functions $q^+(u)$ and $q^-(u)$ analytic in $w\in\mathbb{C}^-$ and
$w\in\mathbb{C}^+$, respectively, such that %
\begin{equation} \label{qprojectiondef}
q=q^++q^-.
\end{equation}
Here we assume that $q(u)\to 0$ for $u\to \pm\infty$.
Eqs.    \e{Projectordef} imply that
\begin{equation} \label{Pfm}
\hat P^+(q^++q^-)=q^+ \quad\text{and}\quad  \hat P^-(q^++q^-)=q^-,
\end{equation}
see more discussion of the operators    \e{Projectordef}  in Ref.
\cite{DyachenkoLushnikovZakharovJFM2019}.

Using Eqs. \e{ComplexPotentialdef}, 
\e{xHy}, \e{PhiTheta1}
and \e{Projectordef} we obtain that
\begin{equation} \label{PiP}
\Pi=\psi+\I \hat H\psi=2\hat P^-\psi
\end{equation}
and
\begin{equation} \label{zP}
z-u=-\hat Hy+\I y=2\I\hat P^-y.
\end{equation}
Analytical continuation of Eqs. \e{PiP} and \e{zP} into the complex
plane $w\in\C$ amounts to straightforward replacing $u$ by $w$ in
the integral representation of  $\hat P^+q(w)$ and $\hat P^-q(w)$ as
detailed in  Ref. \cite{DyachenkoLushnikovZakharovJFM2019}.

Applying  the projector $\hat P^-$ and using Eqs.  \e{PiP},  \e{zP}, one can rewrite (see Ref. \cite{DyachenkoLushnikovZakharovJFM2019}) the dynamical Eqs. \e{implectic2}, \e{Rmatrdef} 
in the complex form

\begin{align}\label{Zt}
&z_t=\I Uz_u, \\
\label{Pit}
&\Pi_t=\I U\Pi_u-B-\mathcal{P,}
\end{align}
where  %
\begin{align} \label{Udef2}
U\equiv\hat P^-(R\bar V+\bar R V)
\end{align}
is the complex transport velocity with %
\begin{equation} \label{Pintdef2}
\mathcal{P}=-\I g(z-w)-2\I\alpha  \hat P^-(Q_u\bar Q-Q\bar Q_u ),
\end{equation}
\begin{equation} \label{Qdef}
Q\equiv\frac{1}{\sqrt{z_u}}=\sqrt{R}
\end{equation}and
\begin{equation} \label{Bintdef2}
B\equiv\hat P^-(|V|^2).
\end{equation}
A complex conjugation $\bar f(w)$ of $f(w)$  in Eqs. \e{Udef2}, \e{Pintdef2}, \e{Bintdef2} and throughout this paper is understood as applied with the assumption that $f(w)$ is the complex-valued function of the real argument $w$ even if $w$ takes the complex values so that %
\begin{equation} \label{bardef}
 \bar f(w)\equiv \overline {f(\bar{w})}.
\end{equation}That definition ensures
  the analytical continuation of $f(w)$ from
the real axis  $w=u$ into the complex plane of $w\in\mathbb{C.}$

Another  equivalent complex form of the dynamical Eqs. \e{implectic2}, \e{Rmatrdef} 
 are the
``Dyachenko'' equations (\citet{Dyachenko2001})

 \begin{align}
\frac{\partial R}{\partial t} &= \I \left(U R_u - R U_u \right), \label{Reqn}\\
\frac{\partial V}{\partial t} &= \I \left[ U V_u - R B_u \right ]+ g(R-1)-2\alpha R  \hat P^-\frac{\p }{\p u}(Q_u\bar Q-Q\bar Q_u ), \label{Veqn}
\end{align}
where %
\begin{align} \label{RVvar1}
&R=\frac{1}{z_u}, \\ \label{RVvar2}
&V=\I\frac{\p \Pi}{\p z}=\I R \Pi_u
\end{align}
are the new unknowns first introduced in Ref.
\citet{Dyachenko2001}. Eqs. \e{Reqn} and \e{Veqn} can be obtained by
differentiating Eqs. \e{Zt}, \e{Pit}  over $u$ and using the
definitions \e{RVvar1} and \e{RVvar2}, see also Ref.
\cite{DyachenkoLushnikovZakharovJFM2019} for more details.

\section{Local analysis: non-persistence of poles in $R$ and $V$ variables and persistence of power law branch points  }
\label{sec:poles}

All four functions  $R$, $V$,  $U$ and $B$ of Eqs. \e{Udef2}, \e{Bintdef2}, \e{Reqn} and \e{Veqn} must have
singularities in the upper half-plane   $w\in\C^+$ while being analytic for   $w\in\C^-$. At the initial time $t=0,$ any singularity for $w\in\C^+$ are allowed including poles, branch points, etc. We are interested in singularities that
keep their nature in the course of evolution to at least a finite duration of time. This ``persistence'' requirement is very restrictive.
It would be extremely attractive to find solutions containing only
pole-type singularities such that   $R$, $V$,  $U$ and $B$
would be the rational functions of $w$. There are examples of different reductions/models of free surface hydrodynamics which allows such rational solutions. They include a free surface  dynamics  for the quantum Kelvin-Helmholtz
instability between two components of superfluid Helium
\citep{LushnikovZubarevPRL2018}; an
interface dynamic between ideal fluid and light highly viscous
 fluid \cite{LushnikovPhysLettA2004}, and
a  motion of  the dielectric fluid with a charged
and ideally conducting free surface in the vertical electric field
\citep{Zubarev_JETPLett_2000,Zubarev_JETP_2002,ZubarevJETP2008eng}.

However, for Dyachenko Eqs. \e{Udef2},\e{Bintdef2}, \e{Reqn} and \e{Veqn}
without surface tension, which takes the following form \begin{align}
\frac{\partial R}{\partial t} &= \I \left(U R_u - R U_u \right), \label{Reqn4}\\
\qquad U&=\hat P^-(R\bar V+\bar RV), \quad B= \hat P^-(|V|^2), \label{UBdef4}\\
\frac{\partial V}{\partial t} &= \I \left[ U V_u - RB_u\right ]+ g(R-1), \label{Veqn4}
\end{align}
rational solutions are not known and we conjecture that they cannot be constructed to satisfy $R(w)\ne 0$ and $|R(w)|<\infty$ for all $w\in \C^-$ (as required by the conformal mapping \e{zwdef} with the condition \e{zwconformal}). The only known exception is the trivial case %
\begin{equation} \label{Vgtrivial}
g=0, \quad \frac{\partial R}{\partial t}\equiv 0, \quad  \text{and} \quad  V\equiv 0,
\end{equation}
 i.e. a stationary solution of fluid at rest without gravity. In that case any singularities (including rational solutions) are allowed in $R(w)$ for   $  w\in\C^+$ and these singularities
 remain constant in time. 
We notice that in Eqs.  \e{Reqn4}-\e{Veqn4} and throughout this paper we use the partial derivatives over $w$ and $u$ interchangeably by assuming the analyticity in $w.$

In this section we provide the local analysis on existence vs. nonexistence  of persistent poles singularities in Eqs.  \e{Reqn4}-\e{Veqn4}. The analysis is local because we use the Laurent series of solutions of free surface hydrodynamics  at any moving point $w=a(t)$, $Im(a)>0$. It means that we are not restricted to rational solutions because such local analysis does not exclude the existence of branch points for $w\ne a(t), \ w\in \C^+$.    In the next section we also provide the local analysis on the persistence of power law branch points.

We note that the conformal map \e{zwdef} and the definition \e{RVvar1} imply that $R(w)\ne0$ for    $w\in\C^-$ and, respectively, %
\begin{equation} \label{Rneq0}
\bar R(w)\ne0 \ \text{for}   \  w\in\C^+.
\end{equation}
 We stress that this is a fact of essential importance. Here and below we often omit the second argument $t$ when we focus on analytical properties in $w$.

 {\it Theorem} 1: Eqs. \e{Reqn4}-\e{Veqn4} have no
persistent in time solution, such that both $R$ and $V$ have only
simple poles singularities at a moving point $w=a(t),$ and a residue
$V_{-1}$ of $V$ is not identically zero in time.

We prove  Theorem 1
``ad absurdum". Simple poles imply that  $V(w)$  and $R(w)$ at $w =
a\in\C^+$ can be written as
\begin{align}\label{Vpole}
V = \frac{V_{-1}}{w - a} + V_{reg}, \quad  \\
R = \frac{R_{-1}}{w - a} + R_{reg}, \quad  \label{Rpole}
\end{align}
where %
\begin{equation} \label{Vreg}
V_{reg}=\sum\limits_{j=0}^{\infty}{V_{j}}{(w - a)^j}
\end{equation}
and
\begin{equation} \label{Rreg}
R_{reg}=\sum\limits_{j=0}^{\infty}{R_{j}}{(w - a)^j}
\end{equation}
 are the regular parts of $V$ and $R$ (these regular parts are the analytic function at
$w=a$). The coefficients $R_j, \ V_{j},\ j=-1,0,\ldots$ and $a$ in
Eqs. \eqref{Vpole}-\e{Rreg} are assumed to be the functions of $t$
only. In a similar way, below we designate by the subscript $``reg"$
the nonsingular part of the all  functions at $w=a. $ The functions
$U(w)$ and $B(w)$ \e{UBdef4} generally also have simple poles at
$w=a$, so that we write them as \begin{align}\label{Upole}
U = \frac{U_{-1}}{w - a} + U_{reg}, \quad U_{reg}=\sum\limits_{j=0}^{\infty}{U_{j}}{(w - a)^j},  \\
B = \frac{B_{-1}}{w - a} + B_{reg}, \quad B_{reg}=\sum\limits_{j=0}^{\infty}{B_{j}}{(w - a)^j}.  \label{Bpole}
\end{align}
To understand validity of these equations and find $U_{-1}$  and $B_{-1}$ we notice that using Eqs. \e{Projectordef}-\e{Pfm} we can rewrite the definitions \e{UBdef4} as
\begin{equation}\label{UBPplus}
\begin{split}
& U=R\bar V+\bar RV-\hat P^+(R\bar V+\bar RV), \\
& B=V\bar V-\hat P^+(V\bar V).
\end{split}
\end{equation}
The functions $\hat P^+(R\bar V+\bar RV)$ and  $\hat P^+(V\bar V)$ are analytic at $w=a\in\C^+$ thus they only contribute to  $U_{reg}$ and $B_{reg},$ respectively. The functions $\bar R$ and $\bar V$ are also analytic at $w=a$ with Taylor series representations
\begin{equation} \label{Rbarexpnasion}
\bar R(w)\equiv R_c+\sum\limits_{j=1}^{\infty}{R_{c,j}}{(w - a)^j}.
\end{equation}
and
\begin{equation} \label{Vbarexpnasion}
\bar V(w)\equiv V_c+\sum\limits_{j=1}^{\infty}{V_{c,j}}{(w - a)^j},
\end{equation}
where   $R_c\equiv\bar R(a)$ and   $V_c\equiv\bar V(a)$ are zero order terms and ${R_{c,j}}$, ${V_{c,j}}$ are the coefficients of the higher order terms of the respective power series.

 Eqs. \e{UBPplus}-\e{Vbarexpnasion} imply that generally $U$ and $B$ have the same types of singularities as $R$ and $V$ except special cases when poles of  either $R$ or $V$  are canceled out.
 Calculating residues of $R\bar V+\bar RV$ and $|V|^2$ at $w=a$  we obtain that
\begin{equation}\label{UBres}
\begin{split}
& U_{-1}=R_cV_{-1}+V_cR_{-1}, \\
& B_{-1}=V_cV_{-1,} \\
& R_c\equiv\bar R(a)\ne0, \quad V_c\equiv\bar V(a),
\end{split}
\end{equation}
where we used Eqs.  \eqref{Vpole}-\e{Bpole}, \e{Rbarexpnasion} and \e{Vbarexpnasion}. Also  $R_c=\bar R(a)\ne0$ follows from the general condition \e{Rneq0}.

According to  Theorem 1's assumption, $V_{-1}\ne 0$. Calculating the
partial derivative of Eq. \e{Vpole},
\begin{align}
\frac{\partial V}{\partial t} &= \frac{a_t V_{-1}}{(w - a)^2} +
\frac{(V_{-1})_t}{w-a} +(V_{reg})_t,
\end{align}
we see that the left-hand side (l.h.s.) of  Eq. \e{Veqn4} has at
most (if $a_t\ne 0$) the second order pole. At the same time, the
right-hand side (r.h.s.) of Eq. \e{Veqn4} has the third order pole
$\frac{-\I R_cV_{-1}^2}{(w-a)^3}$ because $R_c\ne0,$ where we used
Eqs. \e{UBres}. It implies that $V_{-1}=0$ is required to match
l.h.s and r.h.s. of Eq.  \e{Veqn4} which contradicts the initial
assumption thus completing the proof of Theorem 1.

Consider now a more difficult case $R_{-1}\ne 0$ and $V_{-1}=0.$  Then  Eqs.  \eqref{Vpole}-\e{Bpole} and \eqref{Bpole}-\e{UBres} imply that $V(w)$ and $B(w)$ are the regular functions at $w=a.$ If $V_c\ne 0$ then $U(w)$ has a pole according to Eqs. \e{Upole} and \e{UBres}. It leads to the formation of second  order pole in Eq. \e{Reqn4} which is canceled out provided

\begin{equation} \label{at1}
a_t= \I\left[R_0V_c-U_0 \right ],
\end{equation}
where $U_0$ cannot be obtained from the local analysis of this section because it requires to evaluate the projector in Eq. \e{UBPplus} which needs a global information about $V$ and $R$ in the complex plane $w\in \C.$

At the next order, $(w-a)^{-1}$, we obtain that
\begin{equation} \label{wp1}
(R_{-1})_t=-2 \I R_{-1} ( U_1  - R_1V_c )
\end{equation}
and %
\begin{equation} \label{rp1}
B_1={-\I g}{}+V_1V_c,
\end{equation}
where again $U_1$ and $B_1$ can be found only if  $V$ and $R$ are known globally in the complex plane $w\in \C.$
The conditions \e{at1}-\e{rp1} must be satisfied during evolution. Similar  conditions can be obtained from terms of orders  $(w-a)^{0},(w-a)^{1}, \ldots$ to give  equations for time derivatives of coefficients of  the series of regular part of $R$ and $V$ (e.g. the order $(w-a)^{0}$ provides the explicit expressions for $(R_{0})_t$ and $(V_{0})_t$ etc).

 We conclude that the local analysis does not exclude a possibility of the existence of the persistent in time solution with $R_{-1}\ne 0$ and $V_{-1}=0.$  The exceptional case, when the global information is not needed,  is  $V\equiv 0$ (it means that $U\equiv 0$ and $B\equiv 0$) which implies that Eq. \e{rp1} cannot be satisfied for $g\ne 0.$ Then  by contradiction  we conclude that %
\begin{equation} \label{R1mno}
R_{-1}=0, \ \text{for} \ V\equiv 0\ \text{and} \ g\ne 0,
\end{equation}
i.e. no persistent poles exist in that case even for the pole only in $R$ with $V$ analytic at  that point.

Theorem 1 can be generalized to prove nonpersistence of the same higher order poles  $R$ with $V.$ The analysis of that case is beyond the scope of this paper.

We  note that the analysis of Ref. \citet{TanveerProcRoySoc1993}
assumed that both $\Pi_u$ and $z_u$ are analytic in the entire
complex plane $w\in \C$ at $t=0$ (Ref.
\citet{TanveerProcRoySoc1993} actually considered periodic solutions
with an additional symmetry in horizontal direction with the fluid
domain mapped to the unit ball, but we can adjust results of that Ref.
to our conformal map). In terms of $R$ and $V,$ it means that poles
are possible only if $z_u$  has a regular $n$th order zero at $w=a$
with $n=1,2,\ldots$.   Ref.  \citet{TanveerProcRoySoc1993} assumed
$z_u(w=a,t=0)=0$ and $z_{uu}(w=a,t=0)\ne 0$, i.e. $n=1. $ Two cases
were considered in Ref.  \citet{TanveerProcRoySoc1993}  for
$a\in\C^+$: (a) $\Pi_u(w=a,t)\ne 0$ and (b) $\Pi_u(w,t)\equiv 0$ in
$\C$. The case (a) implies that $V_{-1}(w=a,t=0)\ne0$ and
$R_{-1}(w=a,t=0)\ne 0.$ Then our Theorem 1 above proves that
such initial condition cannot lead to persistent pole solutions. It
agrees with the asymptotic result of Ref.
\citet{TanveerProcRoySoc1993}  that a couple of branch points are
formed from that initial conditions during an infinite small duration of  time.  The case (b) of Ref.  \citet{TanveerProcRoySoc1993}
means that $V\equiv 0$ for $t=0$ which has no poles as proven in Eq.
\e{R1mno}.
Refs.   \cite{KuznetsovSpektorZakharovPhysLett1993,KuznetsovSpektorZakharovPRE1994} considered a related case $R\equiv 1$ and a pole in $V$ at $t=0$ which results in the formation of  a couple of branch points  in an infinite small duration of time. That result is again consistent with Theorem 1. Thus our results on the non-existence of persistent
poles are in full agreement with the particular conditions of Refs.
\citet{TanveerProcRoySoc1993,KuznetsovSpektorZakharovPhysLett1993,KuznetsovSpektorZakharovPRE1994}.

We also note that taking into account a nonzero surface tension, i.e. working with Eqs.  \e{Udef2},\e{Bintdef2}, \e{Reqn} and \e{Veqn}
 instead of Eqs. \e{Reqn4}-\e{Veqn4}, immediately shows that pole singularity both for $R$ and $V$ is non-persistent because the dependence of surface tension terms of $Q=\sqrt{R}$  introduces the square root singularity into Eq.  \e{Veqn}
  which cannot be compensated by other terms with poles.

%

Contrary to poles analyzed above,   power law branch points are persistent in time for free surface dynamics which can be shown by the local analysis qualitatively similar to the pole analysis above. The detailed analysis of the persistence of power branch points is however beyond the scope of this paper.     The most common type of branch points, observed in  our
numerical experiments is $\gamma = \frac{1}{2}$ which is consistent
with the results of Refs.
\cite{MalcolmGrantJFM1973LimitingStokes,TanveerProcRoySoc1991,TanveerProcRoySoc1993,KuznetsovSpektorZakharovPhysLett1993,KuznetsovSpektorZakharovPRE1994}.
Square root singularities have been also intensively studied based
on the representation of vortex sheet in Ref.
\cite{MooreProcRSocLond1979,MeironBakerOrszagJFM1982,BakerMeironOrszagJFM1982,KrasnyJFM1986,CaflischOrellanaSIAMJMA1989,CaflischOrellanaSiegelSIAMJAM1990,BakerShelleyJFM1990,ShelleyJFM1992,CaflishEtAlCPAM1993,BakerCaflischSiegelJFM1993,CowleyBakerTanveerJFM1999,BakerXieJFluidMech2011,Zubarev_Kuznetsov_JETP_2014,KarabutZhuravlevaJFM2014,ZubarevKarabutJETPLett2018eng}.

Particular solution of  Eqs. \e{Reqn4}-\e{Veqn4}  is Stokes wave
which is a nonlinear periodic gravity wave propagating with the
constant velocity ~\citep{Stokes1847,Stokes1880}. In the generic
situation, when the singularity of Stokes wave is away from the real
axis (non-limiting Stokes wave), the only allowed singularity in
$\C$ is $\gamma=1/2$ as was proven in Ref.
\citet{TanveerProcRoySoc1991} for the first (physical) sheet  of the
Riemann surface and  in Ref. \citet{LushnikovStokesParIIJFM2016} for
the infinite number of other (non-physical) sheets of Riemann
surface. Refs.
\citet{DyachenkoLushnikovKorotkevichJETPLett2014,DyachenkoLushnikovKorotkevichPartIStudApplMath2016,LushnikovDyachenkoSilantyevProcRoySocA2017}
provided detailed numerical verification of these singularities. The
limiting Stokes wave is the special case   $\gamma=1/3$  with $a=\I
Im(a)$. Also Ref.  \citet{TanveerProcRoySoc1993} suggested the
possibility in exceptional cases of the existence of $\gamma=1/n$
singularities with $n$ being any positive integer as well as
singularities involving logarithms.


\section{New constants of motion for gravity case but with zero surface tension }
\label{sec:Newconstantsmotion}

Assume that both functions $R$ and $V$ are analytic on a Riemann
surface $\Gamma$. The complex plane of $w$ is the first sheet  of
this surface, which we assume to contain a finite number of branch
points $w=w_m, \ m=1,2,\ldots, M$. 

We now address the question if $R$ could have isolated zeros at some other points of $\C^+.$ (We remind that $R(w)\ne 0$ for $w\in\C^-$ because the mapping \e{zwdef}  is conformal.) Assume that $R$ has a simple zero at $w=a(t)$, i.e. $R(a)=0$ and $ R_u(a)\ne 0$. We assume that the functions $R$ and $V$ are analytic at that point witch implies through Eqs. \e{UBPplus} that the functions $U $ and  $B$ \e{UBdef4} are also analytic at that point with  the  Taylor series
\begin{align}\label{Rpoleser2}
&R=\qquad\, {R_1}{(w - a)}+{R_2}{(w - a)^2}+\ldots,\ R_1\ne 0,
\\
&V={V_0+V_1}{(w - a)}+{V_2}{(w - a)^2}+\ldots,  \label{Vpoleser2}
\end{align}
as well as we use the Taylor series
\begin{align}\label{Upoleser2}
&U={U_0+U_1}{(w - a)}+{U_2}{(w - a)^2}+\ldots,
\\
&B={B_0+B_1}{(w - a)}+{B_2}{(w - a)^2}+\ldots.  \label{Bpoleser2}
\end{align}

Similar to Section \ref{sec:poles}, by plugging in Eqs. \e{Rpoleser2}-\e{Bpoleser2} into Eqs. \e{Reqn4} and \e{Veqn4} and collecting terms of the same order of $(w-a)^j$ we obtain for $j=0$ that%
\begin{equation} \label{atzero}
a_t=-\I U_0
\end{equation}
and %
\begin{equation} \label{V0tder}
(V_0)_t=-g.
\end{equation}
The order $j=1$ results in%
\begin{equation} \label{R1der1}
(R_1)_t=0
\end{equation}
and %
\begin{equation} \label{V1der1}
(V_1)_t=g R_1+\I (U_1 V_1-B_1 R_1).
\end{equation}

Equations~\e{V0tder} and \e{R1der1} are of  fundamental
importance. Eq. \e{R1der1} states that both in absence and in presence gravity
\begin{align}\label{R1const}
R_1 = \mbox{const} \equiv\frac{1}{ c_1^{(1)}},
\end{align}
where $c_1^{(1)}$ is the complex time-independent constant.
Eq. ~\e{V0t} results in the trivial dependence on time,%
\begin{equation} \label{V0t}
V_0(t)=-gt+e_1^{(1)},
\end{equation}
where $e_1^{(1)}$ is the complex constant defined by the initial condition,  $e_1^{(1)}=V_0(t=0).$
  Here the subscript $``1"$ stands for the first order of zeros of $R$ in Eq. \e{Rpoleser2}. We conclude that each simple zero of function $R$ generates four
additional real integrals of motion. Two of them are the real and
imaginary parts of $ c_1^{(1)}=1/R_1. $ Two others  are the real and
imaginary parts  of  $e_1^{(1)}=V_0(t)+gt.$ In addition, $V_0(t)$ is
either obeys the trivial linear dependence on time for nonzero
gravity  $g\ne 0$ or coincide with  $e_1^{(1)}$ for   $g=0.$ Eq.
\e{atzero} provides another important relation showing that $-\I
U_0$ is ``the transport velocity" which governs the propagation of
the zeros of the function $R$ in
 the complex plane of $w$.

Taking into account all $N$ isolated simple zeros of $R$ at
$w=a^{(j)}, \ j=1,\ldots, N$ and designating by the superscript
$``(j)"$ the corresponding $j$th zero, we obtain from Eqs.
\e{R1const} and \e{V0t} that $R_1^{(n)} = \mbox{const} \equiv
1/c_1^{(n)}$ and $V_0^{(n)}(t)=-gt+e_1^{(n)}$. Then we notice that
any difference $e_1^{(j)}-e_1^{(n)},\ j,n =1,\ldots N, \ j\ne n,$ is
the true integral of motion even for $g\ne 0.$

We conclude that  $N$ simple isolated zeros of $R,$ separated from
branch points, imply for $g\ne 0$ the   existence of $4N-1$
independent new constants of motion $Re(c_1^{(n)}),$ $Im(c_1^{(n)}),
n=1,\ldots ,N$,   $Re(e_1^{(n)}-e_1^{(N)}),$ $Im(e_1^{(n)}-e_1^{(N)}), \
n=1,\ldots ,N-1$, and $Im(e_1^{(N)})$ as well as one linear function
of time   $Re(V_0^{(N)})=-gt+Re(e_1^{(N)})$. For zero gravity $g=0$ we
have  $4N$ independent new constants of motion $Re(c_1^{(n)}),$
$Im(c_1^{(n)}), Re(e_1^{(n)}),$ $Im(e_1^{(n)}), \ n=1,\ldots, N$.

Section \ref{sec:recoveringsingularities} below demonstrates, in a number of particular cases, the independence  of these  motion constants on time in full nonlinear simulations of Eqs.  \e{Reqn4}-\e{Veqn4}.

Using definitions \e{RVvar1} and \e{RVvar2}, we obtain from Eqs.  \e{Rpoleser2} and \e{Vpoleser2} that %
\begin{equation}\label{zwPiw}
\begin{split}
& z_w=\frac{1}{R_1(w-a)} +(z_w)_{reg},\\
& \Pi_w=\frac{-\I V_0}{R_1(w-a)} +(\Pi_w)_{reg}.
\end{split}
\end{equation}
Eq. \e{R1const}, \e{V0t} and \e{zwPiw} imply that the residues  (i.e. the coefficients  of $(w-a)^{-1}$  of Laurent series),%
\begin{equation} \label{Reszw}
\underset{w=a}{Res}(z_w)=\frac{1}{R_1}=c_1^{(1)} =const
\end{equation}
and
%
$\underset{w=a}{Res}(\Pi_w)=\frac{-\I V_0}{R_1}=-\I c_1^{(1)}e_1^{(1)}=const,
$ 
of both $z_w$ and $\Pi_w$ are constants of motion for $g=0$.  For $g\ne 0,$    $\underset{w=a}{Res}(z_w)$  remains the integral of motion while%
\begin{equation} \label{ResPiwt}
\underset{w=a}{Res}(\Pi_w)=-\I c_1^{(1)}[-gt+e_1^{(n)}],
\end{equation}
i.e. it has the linear dependence on  time.
Section \ref{sec:Anotherveiwmotionconstants} provides another way to straightforward derivation  that these residues are constants of motion.

A Poisson bracket \e{PoissonBracketsRDef} between any motion constant is a motion constant itself (see e.g. Ref. \cite{ArnoldClassicalMechanics1989}). Together such motion constants form a Lie algebra. We conjecture that this Lie algebra is commutative. However, in this paper we are able to prove only a weaker statement that %
\begin{equation} \label{cccommute}
 \{ c_1^{(n)},c_1^{(k)}\}=0
\end{equation}
for any $n,k=1,\ldots,N.$ The proof is almost trivial and relies on the fact that all $ c_1^{(n)}$ integrals are determined by the shape of the free surface $z(u,t)$, ie.e they are functionals of $z$ only. Hence%
\begin{equation} \label{cnvariation}
 \frac{\delta c_1^{(n)}}{\delta \psi}=0, \  n=1,\ldots,N,
\end{equation}
and Eq. \e{cccommute} immediately follows from the Poisson bracket definition \e{PoissonBracketsRDef}.
The question about an explicit calculation of Poisson brackets $\{ c_1^{(n)},e_1^{(k)}\}$ and $\{ e_1^{(n)},e_1^{(k)}\},$  $n,k=1,\ldots,N,$  remains open.

We note that the existence of the arbitrary number of the integrals of motion was not addressed in Ref.  \citet{TanveerProcRoySoc1993} because it focused on the particular case of analytic analytic initial data  in the entire
complex plane $w\in \C.$

\section{Another view of the new motion constants}
\label{sec:Anotherveiwmotionconstants}

In this section we use the dynamical
Eqs. \e{Zt}, \e{Pit} with $\alpha=0.$ It is useful to introduce  new functions %
\begin{equation} \label{rhoWdef}
\rho\equiv z_u=\frac{1}{R} \ \text{and} \ W\equiv
\Pi_u=-\I\frac{V}{R}.
\end{equation}
Then differentiating Eqs. \e{Zt} and \e{Pit} over $u$ together with the definitions \e{rhoWdef} imply that%
\begin{equation}\label{rhoWdynamics}
\begin{split}
& \rho_t=\I(U\rho)_u, \\
& W_t=\I(U W)_u-B_u+\I g(\rho-1).
\end{split}
\end{equation}

Let us address a question about possible singularities of the
functions $\rho$ and $W$. We assume that the functions $R$ and $V$
\e{RVvar1}, \e{RVvar2} have only a  finite number of branch points
for $w\in C^+.$ Apparently, $\rho$ and $W$ generally inherit these
branch points (with the only exception of the possible cancellation
of some branch points because $W=-\I V/R)$  but they cannot have any
additional branch point. In other way, if a branch point appears in
$\rho$ and $W$ at some moment of time, then it immediately implies a
branch point creation in both $R$ and $V.$

However, $\rho$ and $W$ can have poles in the domains of the regularity of $R$ and $V. $ Indeed, assume that $R$ has a regular pole of order $m$ at $w=a$ while $V$ is regular and nonzero at $w=a$. It means that at $w=a$ both $R$ and $V$ can be represented by  Taylor series
\begin{align}\label{Rpoleser4}
&R=\qquad\qquad\qquad\qquad {R_m}{(w - a)^m}+\ldots,\ R_m\ne 0,
\\
&V={V_0+V_1}{(w - a)}+{V_2}{(w - a)^2}+\ldots.  \label{Vpoleser4}
\end{align}
 Then Eqs. \e{rhoWdef},\e{Rpoleser4} and \e{Vpoleser4} imply Laurent series %
\begin{equation}\label{rhoWseries}
\begin{split}
& \rho=\sum\limits^\infty_{j=-m} (w-a)^j\rho_j, \\
& W=\sum\limits^\infty_{j=-m} (w-a)^jW_j .
\end{split}
\end{equation}
Here $\rho_{-1}$ and $W_{-1}$ are the residues or $\rho$ and $W$ at $w=a$ which can be represented by the contour integrals%
\begin{equation} \label{rhom1}
\rho_{-1}=\frac{1}{2 \pi \I}\oint\limits_{C}\rho \,\D w=\frac{1}{2
\pi \I}\oint\limits_{C}\frac{\D w}{R}
\end{equation}
and
\begin{equation} \label{Wm1}
W_{-1}=\frac{1}{2 \pi \I}\oint\limits_{C}W \,\D w=-\frac{1}{2 \pi}\oint\limits_{C}\frac{V\, \D w}{R},
\end{equation}
where $C$ is the counterclockwise closed contour around $w=0$ which
is taken small enough to avoid including any branch point in the
interior.

A direct integration of Eqs. \e{rhoWdynamics} over the contour $C$ implies together with Eqs. \e{rhom1} and \e{Wm1} that %
\begin{align}\label{rhoint}
& \frac{d}{dt}\rho_{-1}=0, \\ \label{Wint} & \frac{d}{dt}W_{-1}=\I g
\rho_{-1},
\end{align}
which is another way to recover the results of Section \ref{sec:Newconstantsmotion} (Eqs. \e{Reszw} and \e{ResPiwt}) in terms of $\rho$ and $W.$ In particular, Eq. \e{rhoint}  means that $\rho_{-1}$ is the constant of motion and $W_{-1}$ is the motion constant only for $g=0$ while generally %
\begin{equation} \label{Wm1t}
W_{-1}=W_{-1}^{(0)}+\I g \rho_{-1}t,
\end{equation}
with $W_{-1}^{(0)}$ being the  complex constant.

Thus poles in $\rho$ and $W$ are persistent in time (at least during a finite time while $w=a$ remains a regular point of both $\rho$ and $W$) which suggests the following decomposition%
\begin{equation}\label{rhoWdecompose}
\begin{split}
& \rho=\rho_{rational}+\rho_b, \\
& W=W_{rational}+W_b,
\end{split}
\end{equation}
where $\rho_{rational}$ and $W_{rational}$ are the rational
functions of $w$ while $\rho_b$ and $W_b$ generally have branch
points.

Assume that at the initial time $t=0$, both $\rho$ and $W$ are
purely rational, i.e. $\rho_b|_{t=0}=   W_b|_{t=0}\equiv 0.$ As a
simple particular case one can assume that
these rational functions have only simple poles with residues $\rho_{-1}^{(k)}$ and $W_{-1}^{(k)}$  at  $N$ points $w=a_k,\ Im(a_k)>0, \ k=1,2,\ldots,N$ as follows %
\begin{equation} \label{rhoWpolesini}
\rho|_{t=0}=1+\sum\limits^N_{k=1}\frac{\rho_{-1}^{(k)}}{w-a_k},
\quad W|_{t=0}=\sum\limits^N_{k=1}\frac{W_{-1}^{(k)}}{w-a_k},
\end{equation}
where 1 in r.h.s of the first equation ensures the correct limit
\e{zlimit}.  Generally these points might be different for  $\rho$
and $W$  but our particular choice of the same points corresponds to
the common poles originating from the zeros of $R$ in Eqs.
\e{rhoWdef}. This type of initial conditions is studied numerically
in Section \ref{sec:recoveringsingularities}. Note that the initial conditions \e{rhoWpolesini}
imply logarithmic singularities at  $w=a_k, \ k=1,2,\ldots,N$  in
both $z$ and $\Pi$ through the definitions \e{rhoWdef} provided $\rho_{-1}^{(k)}\ne 0$ and $W_{-1}^{(k)}\ne 0$.

Bringing Eqs. \e{rhoWpolesini} to the common denominator, we
immediately conclude that $\rho|_{t=0}$ has $N$ zeros (counting
according to their algebraic multiplicity) at some points $w=b_k,\
k=1,2,\ldots, N$. Eq. \e{zwconformal} requires that $Im(b_k)>0$ for
all $ k=1,2,\ldots, N$ which must be taken into account in choosing
initial conditions \e{rhoWpolesini} for simulations.

In a general position $W|_{w=b_k}\ne 0.$ Assume that $w=b_k$ is
$m$th order zero of  $\rho|_{t=0}$.
Then  Eqs. \e{rhoWdef} imply that the Laurent series of both $R$ and $V$ have poles of order $m. $ According to Section \ref{sec:poles} such poles are not persistent in  time meaning that in arbitrary small time they turn into branch points. 

The branch point at $w=b_k$ is generally moving with time, i.e.
$b_k=b_k(t).$ At the initial time $t=0$, the point $w=b_k$ is
separated from all poles $w=a_j, \ j=1,2,.\ldots, N$ in Eqs.
\e{rhoWpolesini}. It means that at least during a finite time
$w=b_k$    will remain separated from  from  poles $w=a_j(t), \
j=1,2,.\ldots, N$ which move according to Eq. \e{atzero} (this equation is also valid for arbitrary $m$ as shown in Section \ref{sec:Newconstantsmotioncapillarity} below for $m=2$, Eq. \e{atzero3}). During
that finite time one can write a decomposition \e{rhoWdecompose} as

\begin{equation}\label{rhoWdecompose2}
\begin{split}
& \rho=1+\sum\limits^m_{j=1}\sum\limits^N_{k=1}\frac{\rho_{-j}^{(k)}}{(w-a_k(t))^j}+\rho_b, \\
& W=\sum\limits^m_{j=1}\sum\limits^N_{k=1}\frac{W_{-j}^{(k)}(t)}{(w-a_k(t))^j}+W_b,
\end{split}
\end{equation}
 where the ``non-rational" terms $\rho_b$ and $W_b$ are identically  zero at $t=0.$ Here $\rho_{-1}^{(k)}$ is the motion constant and $W_{-1}^{(k)}(t)$ is the linear function of time according to Eqs.
\e{rhoint} and \e{Wm1t}. Results of the numerical experiment of
Section \ref{sec:recoveringsingularities} support that decomposition scenario completely.

\section{New constants of motion for nonzero surface tension and second order poles in $z_w$ and $\Pi_w$ }
\label{sec:Newconstantsmotioncapillarity}

If taking into account the nonzero surface tension, $\alpha\ne 0,$ then instead of Eqs.  \e{Reqn4}-\e{Veqn4} we have to consider more general Eqs. \e{Udef2},\e{Bintdef2}, \e{Reqn} and \e{Veqn}.
Expressing $Q=\sqrt{R}$ through $R$ we obtain from Eq. \e{Veqn} that
\begin{align}\label{VQeq2}
\frac{\partial V}{\partial t} = \I \left[ U V_u - R\hat P^- \frac{\p}{\p u}(|V|^2)\right ]+ g(R-1)-\alpha R\hat P^-\frac{\p }{\p u}\left (\frac{R_u\sqrt{\bar R}}{\sqrt{R}}-\frac{\bar R_u\sqrt{ R}}{\sqrt{\bar R}} \right).
\end{align}

Assume that initially $R$ and $V$ satisfy Eqs. \e{Rpoleser2}-\e{Vpoleser2}. Plugging them into r.h.s. of Eq. \e{VQeq2}, one obtains at the leading power of $w-a$ that %
\begin{equation} \label{Vpropo}
\frac{\partial V}{\partial t}\propto\frac{1}{\sqrt{w-a}},
\end{equation}
i.e. a square root singularity appears in $V$ in the infinitely small time. Thus the analysis of Section \ref{sec:Newconstantsmotion} fails for nonzero surface tension. However, we can now consider the double zero in $R$, i.e. Eq. \e{Rpoleser2}
is replaced by %
\begin{align}\label{Rpoleser3}
&R=\qquad\qquad\qquad\qquad{R_2}{(w - a)^2}+\ldots,\ R_2\ne 0,
\\
&V={V_0+V_1}{(w - a)}+{V_2}{(w - a)^2}+\ldots,  \label{Vpoleser3}
\end{align}
and, respectively, the square root disappears in $\sqrt{R}$.

Plugging Eqs.  \e{Upoleser2}, \e{Bpoleser2},\e{Rpoleser3} and \e{Vpoleser3}
into Eqs.  \e{Reqn4} and \e{VQeq2}, and collecting terms of the same order of $(w-a)^j$
we obtain, similar to Section  \ref{sec:Newconstantsmotion},  at order $j=0$ that %
\begin{equation} \label{atzero3}
a_t=-\I U_0
\end{equation}
and %
\begin{equation} \label{V0tder3}
(V_0)_t=-g,
\end{equation}
which are exactly the same  as Eqs. \e{atzero} and \e{V0tder} and which implies that Eq. \e{V0t} is now trivially replaced by %
\begin{equation} \label{V0t2}
V_0(t)=-gt+e_2^{(1)},
\end{equation}
where $e_2^{(1)}$ is the constant defined by the initial condition,  $e_2^{(1)}=V_0(t=0).$
  Here the subscript $``2"$ stands for the second order of zero of $R$ in Eq. \e{Rpoleser3}.

The orders $j=1  $ and $j=2$ result in%
\begin{equation} \label{R2der3}
(R_2)_t=\I R_2U_1
\end{equation}
and %
\begin{equation} \label{V1der3}
(V_1)_t=\I V_1U_1,
\end{equation}
where we do not show an explicit expression for $(V_2)_t$ which appears not much useful.

Excluding $U_1$ from Eqs. \e{R2der3} and \e{V1der3}
we obtain the constant of motion \begin{align}\label{R2V1const}
\frac{V_1}{R_2} = \mbox{const} \equiv f_2^{(1)}.
\end{align}
We note that the surface tension coefficient $\alpha$ does not contribute to Eqs. \e{V0t2} and \e{R2V1const} ($\alpha$ contributes only to
the expression for  $(V_2)_t$ and higher orders in powers of $w-a)$.
Thus Eq.  \e{R2V1const} is valid for arbitrary $g$ and $\alpha$.

Eqs.  \e{RVvar1},\e{RVvar2},\e{Rpoleser4} and \e{Vpoleser4} imply that Eq. \e{zwPiw} is replaced by  Laurent series
\begin{equation}\label{zwPiwn}
\begin{split}
& z_w=\frac{1}{R_2(w-a)^2} -\frac{R_3}{R_2^2(w-a)}+O\left((w-a)^{0}\right),\\
& \Pi_w=\frac{-\I V_0}{R_2(w-a)^2} +\frac{\I(R_3 V_0-R_2V_1)}{R_2^2(w-a)}+O\left((w-a)^{0}\right).
\end{split}
\end{equation}

However, Eqs. \e{V0tder3} and \e{R2V1const} do not exhaust all
integrals of motion for the case of the second order pole of this
section.  For that we note that the results of Section
\ref{sec:Anotherveiwmotionconstants} on time independence of
$\rho_{-1}=\underset{w=a}{Res}(z_w)$ and linear dependence of
$W_{-1}=\underset{w=a}{Res}(\Pi_w)$   on time (see Eqs. \e{rhoint}
and \e{Wint}) are true for  the second order pole as well as they
remain valid for $\alpha\ne0$. We note that it is possible to derive
Eqs. \e{rhoint} and \e{Wint} by direct computations in $R$ and $V$
variables, similar to the  derivation of Eq. \e{R2V1const}, but we
do not provide it here because the analysis of  Section
\ref{sec:Anotherveiwmotionconstants} is much more elegant for these
residues. Thus Eq. \e{rhoint} imply that it is natural to replace
the  definition of the motion constant $ c_1^{(1)}$  from Eq.
\e{Reszw} of Section   \ref{sec:Newconstantsmotion}    for the first
order pole by
\begin{equation} \label{Reszwm}
\underset{w=a}{Res}(z_w)\equiv c_2^{(1)} =const
\end{equation}
for the second order pole, where the subscript ``2" means that second order. Using Eq. \e{V0t2}, one can also rewrite Eq. \e{Wm1t} as follows %
\begin{equation} \label{PiuRest}
\underset{w=a}{Res}(\Pi_w)+\I\underset{w=a}{Res}(z_w)V_0=const.
\end{equation}

  The explicit expressions for   $\underset{w=a}{Res}(z_w)$ and  $\underset{w=a}{Res}(\Pi_w)$    immediately follow from Eq. \e{zwPiwn} giving that
 \begin{align}\label{ReszwPiw2}
&
 \underset{w=a}{Res}(z_w)=\frac{-R_3}{R_2^2} ,\quad 
\underset{w=a}{Res}(\Pi_w)=-\I\underset{w=a}{Res}(z_w)V_0-\frac{\I
V_1}{R_2}. 
%
%
\end{align}
Eqs. \e{R2V1const} and \e{ReszwPiw2} also imply that Eq. \e{PiuRest}  is not the independent integral of motion.

We now generalize the statement of Section
\ref{sec:Newconstantsmotion}  on the number of motion constants  to
the   second order pole case of this section. Taking into account
all $N$ isolated  zeros of the second order  of $R$ at $w=a^{(j)}, \
j=1,\ldots, N$ and designating by the superscript $``(j)"$ the
corresponding $j$th zero, we  obtain $2N$ real independent integrals
of motion   $Re(f_2^{(j)}),Im(f_2^{(j)}),j=1,\ldots,N$   from  Eq.
\e{R2V1const}; $2N-1$  real independent integrals of motion
$Re(e_2^{(j)}-e_2^{(N)}), Im(e_2^{(j)}-e_2^{(N)}),\ j =1,\ldots
N-1,$ $Im(e_2^{(N)})$ as well as one linear function of time
$Re(V_0^{(N)})=-gt+Re(e_2^{(N)})$  (similar to Section
\ref{sec:Newconstantsmotion}, that number of integrals turns into $2N$ for $g=0$
by adding $Re(e_2^{(N)})$) from Eq. \e{V0t2}; and $2N$ real
independent integrals of motion
$Re(c_2^{(j)}),Im(c_2^{(j)}),j=1,\ldots,N$   from Eq. \e{Reszwm}.
Thus the total number of independent complex integrals of motion is
either  $6N-1$ for $g\ne 0$  or $6N$ for $g=0$.  All these results
for the motion constants are valid for nonzero surface tension
$\alpha\ne 0.$  We note that if we look at the poles of the order
higher than two, the number of the independent integrals of motion
is increasing (with $\alpha\ne 0$ allowed for all even orders).
However such general case of  third and higher order poles is beyond
the scope of this paper.

One can easily generalize the results of both this section and Section   \ref{sec:Newconstantsmotion}   by allowing   a mixture of the terms with the highest first and the second order poles (correspond to Eqs.  \e{Rpoleser2} and \e{Rpoleser3}, respectively) at each of $N$ points of zero of $R.$ The corresponding number of the independent integral of motion can be easily recalculated for that more general case.

The constants of motion $c_2^{(j)},j=1,\ldots,N$   \e{Reszwm} are   functionals of $z$ only, similar to $c_1^{(j)},j=1,\ldots,N$ of Section        \ref{sec:Newconstantsmotion}. It implies an immediate  generalization of  Eqs. \e{cccommute} and \e{cnvariation} to  %
\begin{equation} \label{cccommutem}
 \{ c^{(n)}_{m_1},c^{(k)}_{m_2}\}=0
\end{equation}
 and%
\begin{equation} \label{cnvariationm}
 \frac{\delta c_m^{(n)}}{\delta \psi}=0
\end{equation}
for any $n,k=1,\ldots,N$  and any $1\le m,m_1,m_2\le 2.$

 \section{Kelvin theorem for phantom hydrodynamics and global analysis }
\label{sec:Kelvintheorem}

In this section we return to the analysis of the free surface
hydrodynamics in terms of the functions $R$ \e{RVvar1} and $V$
\e{RVvar2} which satisfy the Dyachenko Eqs. \e{Udef2}, \e{Qdef},
\e{Bintdef2}, \e{Reqn} and \e{Veqn}. Similar to Section
\ref{sec:Anotherveiwmotionconstants}, we  assume that both $R$ and
$V$   have a finite number of branch points and pole singularities
for $w\in\C^+$. As discussed in Section \ref{sec:poles},
 Eqs. \e{UBPplus}-\e{Vbarexpnasion} imply that generally the functions $U$ and $B$ have the same types of singularities as $R$ and $V$ except special cases of cancelation of singularities. Moreover, both  $U$ and $B$ can have singularities only at points were  $R$ and $V$  also have singularities. Also all four functions $R$, $V$, $U$ and $B$ are analytic for $w\in\C^-$.
We note that beyond branch point our analysis cannot fully exclude the appearance of essential singularities. However, all our numerical simulations of Section \ref{sec:recoveringsingularities} indicate only a formation of branch points which is also consistent with the assumption of this section that the only possible singularities of are poles and branch points. See also Ref. \citet{LushnikovStokesParIIJFM2016}  for similar discussion in the particular case of Stokes wave.

We stress that the main task of the theory is to address the
analytic properties of  $R$ and $V$ in the entire complex plane
$w\in \C$.   Moreover, we consider an analytical continuation of
these functions into the Riemann surfaces which we call by
$\Gamma_R(w)$ and  $\Gamma_V(w)$, respectively. It means that we
need a global analysis beyond the local analysis of Sections
\ref{sec:poles}-\ref{sec:Anotherveiwmotionconstants}. Little we know
about these surfaces. If either $R$ or $V$ would be a purely
rational function, then the corresponding Riemann surface would have
a genus zero (see e.g. Ref.
\citep{DubrovinFomenkoNovikovBookPartII1985}. However, the results
of Section \ref{sec:poles} suggest that such rational solutions are
unlikely to exist for any finite duration of time. The local
analysis of Section \ref{sec:poles}  suggests that a branch point in
$V$ implies that $R$ also has a branch point of the same type at
that point. Then we expect that a covering map exists from
$\Gamma_R(w)$ onto  $ \Gamma_V(w).$
 Then from Eq. \e{UBPplus} we conclude that $U$  has the same Riemann surface as    $\Gamma_R(w)$  while $B$ has the same Riemann surface as    $\Gamma_V(w).$ We conjecture that in the general case,  branch points of  $\Gamma_R(w)$ and  $ \Gamma_V(w)$ are of square root type, i.e. their genuses are nonzero.
We also conjecture, based on results of Section    \ref{sec:poles},   that $V(w)$ generally has no poles for $w\in \C$  with the same  valid for $B(w$).
We conjecture that in a general position   $\Gamma_R(w)$ and  $ \Gamma_V(w)$ are non-compact surfaces with the unknown total number of
sheets. Our experience with the  Stokes wave
\citep{LushnikovStokesParIIJFM2016} suggests that generally the
number of sheets is infinite. Some exceptional cases like found in
Refs. \cite{KarabutZhuravlevaJFM2014,ZubarevKarabutJETPLett2018eng}
have only a finite number of sheets of Riemann surface (these
solutions however have diverging values of $V$ and $R$ at
$w\to\infty$). We suggest that the detailed study of such many- and
infinite-sheet Riemann surfaces is one of the most urgent goal in
free surface hydrodynamics. This topic is however beyond the scope
of this paper. 

Both Riemann surfaces    $\Gamma_V(w)$ and $\Gamma_R(w)$ appear {\it
after} we define the conformal mapping \e{zwdef}. There is another
Riemann surface, which we call by $G(z),$ appearing {\it before} the
conformal mapping. Indeed, we can look at the complex velocity $V$
inside the fluid's in the  complex $z$ plane 
using the
definition \e{RVvar2}. An analytical continuation of $V(z,t)$  to
outside of fluid defines $G(z)$. For stationary waves such
continuation had been considered since 19th century, see e.g. Ref.
\citet{Lambbook1945}. $\Gamma_V(w)$ is the composition of $G(z)$ and
$z(w)$ as $\Gamma(w)=G(z(w))$. The analytical continuation of the
time-dependent Bernoulli Eq. \e{dynamic1} also allows to recover a
fluid pressure  in $z$ plane.

The analytically continued function $V(z,t)$ describes a flow of the
imaginary (fictional) fluid on the Riemann surface $G(z)$. We call
the corresponding theory   ``the phantom hydrodynamics". We
introduce that new concept in effort to find a physical
interpretation of new motion integrals found in Sections
\ref{sec:Newconstantsmotion}-\ref{sec:Anotherveiwmotionconstants}.
The idea of using the circulation over complex contour in the domain
of analyticity of the analytical extension  as the integral of
motion was also introduced \cite{CrowdySIAMJApplMath2002} in quite
different physical settings of  the rotating Hele-Shaw problem and
the viscous sintering problem. For Hele-Shaw problem (in the
approximation of the Laplace growth equation) the infinite number of
the integrals of motion were also discovered in
 Ref. \cite{RichardsonJFM1972} and  later used in Ref. \citep{MineevWeinsteinWiegmannZabrodinPRL2000}  to show the integrability of that equation in a sense of the existence of  infinite number of
integrals of motion and its relation to the dispersionless limit of
the integrable Toda hierarchy.

Thereafter we  assume that the non-persistence of poles is valid for
any order of poles both in $V$ and $R$ (as was proved for more
restricted cases in Theorem 1 and 2 of Section    \ref{sec:poles}
(it also means that we fully exclude a trivial case given by Eq.
\e{Vgtrivial}). Then $R$ can be analytically continued to the same
surface as $\Gamma_V$ without the introduction of  additional
singularities, i.e.  $\Gamma_R= \Gamma_V$. Respectively, one can consider
both Eqs. \e{rhom1} and \e{Wm1} on the whole surface $\Gamma_V$
beyond just  $w\in \C.$ Now $C$ in Eqs. \e{rhom1} and \e{Wm1} is any
closed and small enough contour on $\Gamma_V$, which moves together
with the surface. It means that poles of both $\rho$ and $W$ on
other sheets of  $\Gamma_V$ generate integrals of motion and the
total number of these integrals is unknown. One can consider these
integrals on the physical surface $G.$ As far as $\frac{d w}{R}=dz$,
one can rewrite r.h.s. of  Eq. \e{Wm1} as $-\frac{1}{2 \pi}\oint\limits_{}V\D z$ and interpret a conservation of $W_{-1}$ at $g=0$ as a
``generalized Kelvin theorem" valid for the phantom hydrodynamics
(see e.g. \citet{LandauLifshitzHydrodynamics1989} for the Kelvin
theorem of the usual hydrodynamics). Notice however, that this
generalized Kelvin theorem can be formulated only {\it after}  the
conformal mapping of the surface $G$ to the surface $\Gamma_V.$

\section{Numerical simulations of free surface hydrodynamics through the additional time dependent conformal mapping}
\label{sec:numericalsimulations}

\subsection{Basic equations for simulations and spectrally accurate adaptive mesh refinement }
\label{subsec:simulationcode}

We performed simulations of Dyachenko Eqs. \e{Udef2}, \e{Qdef},
\e{Bintdef2}, \e{Reqn} and \e{Veqn} using pseudo-spectral numerical
method based on Fast Fourier transform (FFT) coupled with an
additional conformal mapping (Ref.
\cite{LushnikovDyachenkoSilantyevProcRoySocA2017})
\begin{equation} \label{qnewdef}
q=q^*+2\,\text{arctan}{\left [ \frac{1}{L}\tan{\frac{w-w^*}{2}}  \right]}.
\end{equation}
Eq.  \e{qnewdef} provides the mapping from our standard conformal
variable $w=u\ I v$  into the new conformal variable $q$. Here $L,
q^*,w^*\in\R$ are the parameters of  that additional conformal
mapping. The details of the  numerical method are provided in  Ref.
\cite{LushnikovDyachenkoSilantyevProcRoySocA2017}.   Here and below
we assume without loss of generality that both $R$ and $V$ are the
periodic functions of $w$ with the period $2\pi$ (if the period
would be different then one can rescale independent variables $w$
and $t$ as well as $g$ and $\alpha$ to ensure  $2\pi$ periodicity
while keeping the same form of Eqs. \e{Udef2},\e{Bintdef2}, \e{Reqn}
and \e{Veqn}). To recover the limit of decaying solution at
$|u|\to\infty$ considered in previous sections, we take the limit of
large spatial period (before rescaling to $2\pi$). In terms of
rescaled variables, it means that the distance of complex
singularities of interest to the real line $u=w$ must be much
smaller than $2\pi$. However the analytical results of previous
sections are valid for the periodic case also. See also Ref.
\cite{DyachenkoLushnikovKorotkevichPartIStudApplMath2016}) for  the
detailed discussion of the periodic case compared with the decaying
case. We also note that the conformal map \e{qnewdef} conserves
$2\pi$ periodicity of both $R$ and $V$.

The goal of our simulations was to reach a high and a well-controlled numerical precision while maintaining the analytical properties in the complex plane. The reason of using the new conformal variable \e{qnewdef} for simulations    is that a straightforward representation of $R$ and $V$ by Fourier series (while ensuring the analyticity of both $R$ and $V$ for $w\in \R$) would
turn much less efficient as  the lowest  complex singularity at $w=w_c\equiv u_c+\I  v_c$ of  $R$ or/and $V$ approaches the real line during dynamics. Such approach would imply a slow decay of the Fourier coefficients
as%
\begin{equation} \label{expspectral}
 \propto e^{-v_c |k|} \quad  \text{for} \quad k\gg 1,
\end{equation}
 where $k$ is the Fourier wavenumber.
It was found in Ref. \cite{LushnikovDyachenkoSilantyevProcRoySocA2017} that the  conformal mapping \e{qnewdef}  allows to move the singularity $w=w_c$ significantly away from the real line. It was shown in that Ref. that the optimal choice of the parameter $L$ is
\begin{equation} \label{Loptimal}
L=L_{optimal}\simeq \left (\frac{v_c}{2} \right )^{1/2}
\end{equation}
which ensures a mapping of $w=w_c$ into $q=q_c,$  $Im(q_c)\approx(2v_c)^{1/2}\gg  v_c$ for $v_c\ll 1$   and the fastest possible convergence of Fourier modes in $q$ variable as
\begin{equation} \label{vc1p2}
\propto e^{-(2v_c)^{1/2} |k|} \quad \text{for} \quad k\gg 1.
\end{equation}
The parameters $u^*$ and $q^*$ of Eq. \e{qnewdef} are  $u^*=u_c$ and
$q^* = 2\arctan[L\tan{({u^*}/{2})}]$. The introduction of these
parameters is a modification of the results of    Ref.
\cite{LushnikovDyachenkoSilantyevProcRoySocA2017}  to account for
the motion of complex singularities in the horizontal direction. The
scaling \e{vc1p2} is greatly beneficial compared with the scaling
\e{expspectral} for $v_c\ll1$ because to reach the same numerical
precision one needs to take into account a factor
$\sim(v_c/2)^{1/2}$ less  Fourier modes. E.g., Ref.
\cite{LushnikovDyachenkoSilantyevProcRoySocA2017} demonstrated $\sim
10^6$ fold speed up  of simulations of Stokes wave with $v_c\simeq
10^{-11}.$ In our time-dependent simulations described below we
routinely reached  down to $v_c\simeq 10^{-6}$. It is definitely
possible to extend our simulations for significantly smaller $v_c$
which is however beyond the scope of this paper which is focused on
numerical verifications of analytical results of above sections.

Our simulation method is based on the representation of $R$ and $V$ in Fourier series in $q$ variable as $R(q,t) = \sum\limits_{k= 0}^{-\infty} R_k(t) e^{\I kq}$ and $
V(q) = \sum\limits_{k=0} ^{-\infty}V_k (t)e^{\I kq}$, where $R_k(t)$ and $V_k(t)$ are Fourier modes for the integer wavenumber $k.$ These modes are set to zero for $k>0$ which ensures analyticity of $R$ and $V$ for $w\in\C^-. $  For simulations we truncated Fourier series to the finite sums $R(q,t) = \sum\limits_{k= 0}^{-N} R_k(t) e^{\I kq}$ and $
V(q) = \sum\limits_{k=0} ^{-N}V_k (t)e^{\I kq}$,  where the integer $N$ is a time dependent and chosen large enough at each $t$ to ensure about round-off double precision $\sim 10^{-16}$. Eqs. \e{Udef2},\e{Bintdef2}, \e{Reqn} and \e{Veqn} were rewritten in $q$ variable with the main difficulty to numerically calculate the projector $\hat P^-$ (defined by Eq. \e{Projectordef} in $u$ variable but must be numerically calculated in $q$ variable) which we  did based on Ref. \cite{LushnikovDyachenkoSilantyevProcRoySocA2017}. We used  the uniform grid in $q$ for FFT which we call  the computational domain.   In $u$ variable such grid  implies a highly non-uniform grid which focuses on the domain closest to the lowest complex singularity, see  Ref. \cite{LushnikovDyachenkoSilantyevProcRoySocA2017} for details. In other words, our numerical method provides a spectrally accurate adaptive mesh refinement.

During dynamics we fixed  $N$, $u^*$ and  $L$ for a finite period of
time during which Fourier spectrum was resolved up to   prescribed
tolerance (typically we choose that tolerance $\sim 10^{-13}$ for
the double precision simulations). The advancing in time was
achieved by the six order Runge-Kutta  method with the adaptive time
step to  both maintain the numerical precision and satisfy the
numerical stability. The   de-aliasing  (see e.g. Ref.
\cite{BoydChebyshevFourierSpectralMethodsBook2001}) was not required
because after each time step we set all positive Fourier modes to
zero to ensure  analyticity for $w\in\C^-.$  If Fourier spectrum at
some moment of time  turned too wide to meet the tolerance (this
occurs due to the motion of the lowest singularity $w_c$ in $\C^+$)
then we first attempted to adjust  $u^*$ and  $L$  to make the
spectrum narrower to meet the tolerance. This is achieved through
the approximation of   $u_c=u^*$ by the location of the maximum of
the Jacobian ${|z_u|^2}$   at the real line $w=u$ while  the updated value of $L$
was obtained by decreasing $L$ by a factor $2^{1/2}.$
Alternative procedure to find more accurate values of $v_c$ (and respectively more accurate value of $L$ through Eq. \e{qnewdef}) is
to either use the asymptotic of Fourier series as in Refs.
\citet{DyachenkoLushnikovKorotkevichJETPLett2014,DyachenkoLushnikovKorotkevichPartIStudApplMath2016}
or perform the least-square-based rational approximation  of solution
(described below) to find an updated value of $w_c$ and,
respectively to update $u^*$ and $L$. After finding a new values of
$u^*$ and $L$, the spectral interpolation was performed to the new
grid with the updated  values $u^*$ and $L$. That step cannot be
performed with FFT because the change of   $u^*$ and $L$ causes a
nonlinear distortion of the uniform grid compared with the previous
value of $L$. Instead, straightforward evaluations of Fourier series
at each new value of $q$ were performed  requiring $\sim N^2$ flops
(while FFT requires only $\sim N\log N$ flops). However, such change
of $L$ and/or $u^*$ was required typically once a few hundreds or
even many thousands of time steps so the added numerical cost from
that $N^2$ flops step was moderate. If such first  attempt to update
$u^*$ and $L$ was not sufficient to meet the tolerance, $N$ was also
additionally increased by the spectral interpolation to the new grid
in $q$ by adding extra  zeroth Fourier modes (i.e. increasing $N$)
and calculating numerical values on the new grid through FFT.

\subsection{Recovering of motion of singularities for $w\in \C^+$ by  the  least square rational approximation }
\label{sec:rationalapproximation}

The simulation approaches of Section \ref{subsec:simulationcode}
results in the numerical approximation of $R$ and $V$ on the real
line $w=u$ for each $t$. To recover the structure of complex
singularities of $R$ and $V$ for $w\in \C^+$ for each $t$ we used
the  least-square rational approximation  based on the
Alpert-Greengard-Hagstrom (AGH) algorithm ~\cite{AGH2000} adapted to
water waves simulations in
Ref.~\cite{DyachenkoLushnikovKorotkevichPartIStudApplMath2016}.
Contrary to the analytical continuation of Fourier series (see e.g.
\citet{DyachenkoLushnikovKorotkevichJETPLett2014,DyachenkoLushnikovKorotkevichPartIStudApplMath2016}),
AGH algorithm allows the analytical continuation from   the real
line $w=u$ into $w\in\C^+$ well above the lowest singularity
$w=w_c.$ AGH algorithm is based on approximation of the function
$f(u)$ with the function values given on the real line $w=u$ by the
rational function in the least square sense. The rational
approximant is then straightforward to analytically continue to the
complex plane by replacing $u$ by $w$. AGH algorithm overcomes
numerical instabilities typical for  Pad\'e approximation  (see e.g
\cite{BakerGravesMorrisBook1996}) which is based on value of
function and its derivative in a single point, see Refs.
\citet{GonnetPachonTrefethenElectrTransNumAnal2011,DyachenkoLushnikovKorotkevichPartIStudApplMath2016}
for more discussion. AGH algorithm robustly recovers poles in
solution while branch cuts are approximated by a set of poles as
follows

\begin{equation}
\label{approx_cut} g(\zeta)=\frac{1}{2\pi} \int\limits_{C}
\dfrac{\rho(\zeta')\D\zeta'}{\zeta -\zeta'} \simeq  \sum\limits_{n =
1}^{N} \dfrac{\sigma_n}{\zeta -\zeta_n},
\end{equation}
where the function $g(\zeta)$ has s single branch cut along the
contour $C$ in the complex plane of $\zeta$ with the $\rho(\zeta)$
being a jump of $g(\zeta)$ at the branch cut. R.h.s. of Eq.
\e{approx_cut} approximates $g(\zeta)$ by simple poles located at
$\zeta=\zeta_n\in C, \ n=1,\ldots, N$ with    the residues
$\sigma_n, \ n=1,\ldots, N$. A generalization to multiple branch cut
is straightforward. Ref.
\citet{DyachenkoLushnikovKorotkevichPartIStudApplMath2016}
demonstrated for the particular case of Stokes wave that
$\rho(\zeta)$ can be robustly recovered from $\zeta_n$ and
$\sigma_n$ by increasing $N$ with the increase of the numerical
precision. For fixed $N$, r.h.s. of Eq. \e{approx_cut} approximates
$g(\zeta)$ with high precision for all points $\zeta\in \C$ located
away from $C$ by a distance several times exceeding the distance
between neighboring $\zeta_n.$  In numerical examples below we
distinguish actual poles of $g(w)$ from the artificial poles which
occur in approximation of branch cuts, as in Eq. \e{approx_cut}, by
changing the numerical precision (the actual poles remains the same
while the number of poles in approximation \e{approx_cut} increases
with the increase of the numerical precision). Alternative way is to
look at the dynamics of poles: while actual poles move continuously
with time and their residues either remain constant or change
gradually in time (in accordance with the analysis of Sections
\ref{sec:Newconstantsmotion}-\ref{sec:Newconstantsmotioncapillarity}),
the poles approximating branch cuts quickly change both their
positions and residues with their number $N$ also  changing as seen
in numerical examples of Section \ref{sec:recoveringsingularities}
below.

To take into account $2\pi$ periodicity of our simulation in $w$ variable we define an auxiliary conformal transformation  %
\begin{equation} \label{zetadef}
\zeta = \tan\frac{w}{2}
\end{equation}
which maps the stripe  $-\upi <Re(w)<\upi$ into  the complex $\zeta$
plane. Also $w\in\C^+(\C^-)$ imply that $\zeta\in\C^+(\C^-)$, see
also Ref.
\citet{DyachenkoLushnikovKorotkevichPartIStudApplMath2016} on more
details of the mapping \e{zetadef}.  $\zeta$ variable is convenient
to use in AGH algorithm
 (\citet{DyachenkoLushnikovKorotkevichPartIStudApplMath2016}) which is assumed below.

 While the simulations of dynamics were performed in  double
precision arithmetic,  AGH algorithm was  performed in variable
precision   (typically we used 512 bits, i.e. approximately 128
digits). It is also possible to use a variable precision for
dynamics (as was done in Ref.
\citet{DyachenkoLushnikovKorotkevichPartIStudApplMath2016} for Stoke
wave) to improve a numerical approximation of branch cuts which is
however beyond the scope of this paper.


\section{Recovering a motion of singularities from simulations and comparison with analytical results}
\label{sec:recoveringsingularities}


The initial data for $z_u$ and $\Pi_u$ (which immediately implies the initial data for $R$ and $V$ through the definitions \e{RVvar1} and \e{RVvar2}) were chosen in the rational form for the variable $\zeta$ \e{zetadef} which ensures $2\pi$ periodicity in $w$ variable.
Below we count a number poles per  period, i.e. inside a single  stripe  $-\upi <Re(w)<\upi$ which is the same number as in the complex plane of $\zeta.$

\subsection{A pair of simple poles in initial conditions and a formation of oblique jet}
\label{sec:simplepoles}



Consider an initial condition in the form of a pair of simple poles at $w=a_1(0)$ and $w=a_2(0)$ both for  $z_u$ and $\Pi_u$ as follows
 \begin{equation}\label{ini1}
 \begin{split}
&z_u= 1 - q\left[ \cot\left( \frac{w - a_1(0)}{2} \right) - \cot\left( \frac{w - a_2(0)}{2} \right) \right]\\
&\quad=1 - q\left[ \frac{1+\zeta\tan{\frac{a_1(0)}{2}}}{\zeta-\tan{\frac{a_1(0)}{2}}}- \frac{1+\zeta\tan{\frac{a_2(0)}{2}}}{\zeta-\tan{\frac{a_2(0)}{2}}}\right], \\
&\Pi_u = \I c\left(1 - z_u \right),
\end{split}
\end{equation}
where $a_1(0),a_2(0)\in\C^+,$  $c,q\in\C$ are constants and  we used the trigonometric identity
\begin{align} \label{tansum}
\cot{(a-b)}=\frac{1+\tan{a}\tan{b}}{\tan{a}-\tan{b}}. 
\end{align}

Eqs. \e{ini1} and \e{RVvar1} and \e{RVvar2} imply that $R=1/z_u$ is analytic and has simple zeros   at  $w=a_1$ and $w=a_2$ while $V$ is analytic and nonzero at these points provided $q\ne 0$ and $c\ne 0$ which corresponds to the case of Eqs. \e{Rpoleser2} and \e{Vpoleser2}.

The conformal map \e{zwdef} requires that $z_u\ne 0$ for $w\in\C^-$. Solving for $z_u=0$ in the first Eq. of  \e{ini1} results in %
\begin{align} \label{wzero1}
w_{\pm}=2\arctan\left [\frac{{A_1}+{A_2}\pm\sqrt{\left(1-4
q^2 )(A_2-A_1)^2+4 q (A_1 A_2+1)(A_1-A_2) \right)}}{2(1- q[A_1 -A_2
])}\right ], \nonumber \\
A_1\equiv\tan{\frac{a_1(0)}{2}}, \quad A_2\equiv\tan{\frac{a_2(0)}{2}}
\end{align}
which provides a restriction on allowed numerical values of $q,a_1(0)$ and $a_2(0)$ to ensure that $w_\pm\in\C^+.$

We choose %
\begin{equation} \label{inita2a2p01}
 a _1(0)= 0.3\I, \quad a_2 (0)= 0.6\I, \quad c=0.64/q \quad \text{and} \quad  q = 0.4\exp\left(\frac{3\pi}{5}i\right).
\end{equation}
Eqs. \e{wzero1} and \e{inita2a2p01} result in%
\begin{equation} \label{wpm01}
w_+=0.465388\ldots + \I \,0.532846 \ldots \quad \text{and} \quad  w_-=-0.465388 \ldots+ \I\,0.367154\ldots, \end{equation}
i.e. $w_\pm\in\C^+$ in this case as required.
Taylor series expansions of $z_u$ and $\Pi_u$  \e{ini1} at
 $w=w_\pm\in\C^+
$ and $t=0$ reproduce Eqs. \e{Vpole} and \e{Rpole} in the variables
$R$ and $V$ with $R_{-1}\ne 0$ and $V_{-1}\ne 0$. Then Theorem 1 of
Section \ref{sec:poles} proves that solutions  \e{Vpole} and
\e{Rpole} are not persistent in time. Generally we expect a
formation of a pair of square root branch points from $w=w_\pm$ at
arbitrary small time $t>0$ which is also consistent with Refs.
\cite{TanveerProcRoySoc1993,KuznetsovSpektorZakharovPhysLett1993,KuznetsovSpektorZakharovPRE1994}.
 The initial poles at $w=a_1(0)$ and $w=a_2(0)$  are expected to be
persistent for at least a finite time duration according to the
results of Section \ref{sec:Newconstantsmotion}.

\begin{figure}
\includegraphics[width=0.495\textwidth]{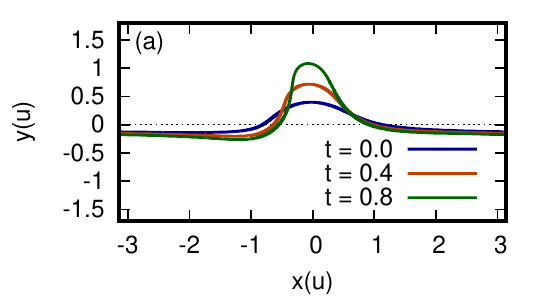}
\includegraphics[width=0.495\textwidth]{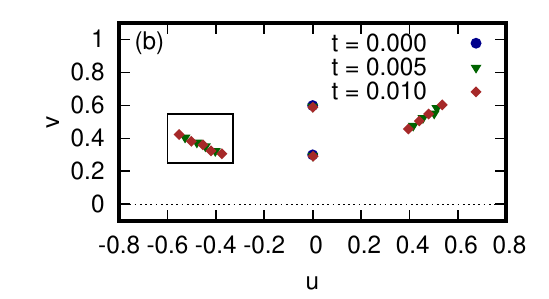}
\includegraphics[width=0.495\textwidth]{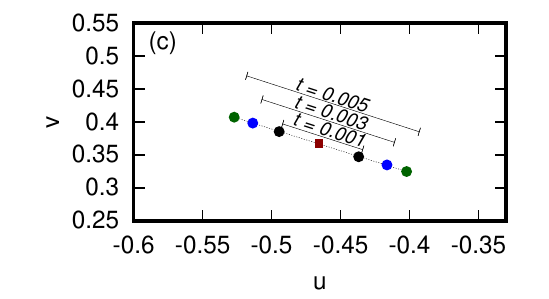}
\includegraphics[width=0.495\textwidth]{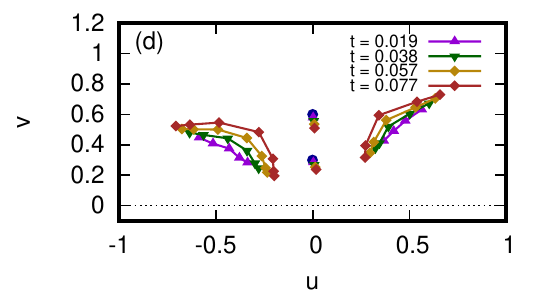}
\caption{ Simulations with the initial conditions
\e{ini1},\e{inita2a2p01} and $g=\alpha=0.$ (a) Profiles of free
surface at different times. (b) Complex singularities of $z_u$
recovered by AGH algorithm at  small times including the initial
time $t=0.$ Two persistent poles recovered by AGH algorithm are
shown by thick dots of different style moving near the imaginary
axis. It is seen that these two poles,  originating from  the
initial conditions (their initial  positions are exactly at the
imaginary axis according to Eq.  \e{inita2a2p01}), only slightly
move away from the initial positions at these early times. Two
initial zeros of $z_u$ located at $w=w_\pm$  according to Eq.
 \e{wpm01} turn into two short branch cuts at
arbitrary short times. Each branch cut connects two branch points.
These branch cuts are revealed in AGH algorithm  by a dense set of
poles located near  $w=w_\pm$   with the number of these poles
growing with time. (c) The schematic zoom into a small area around
$w=w_-$ (in (b) that area is shown by the rectangular frame around
the left branch cut) to display the extension of branch cut with
time. The small filled square shows   the point $w=w_-$. The length
of each branch cut grows approximately linearly with time.   (d) The
same as in  (b) but at larger times when the length of branch cuts
reaches $\sim 1.$ Poles approximating branch cuts are connected by
solid lines.  } \label{e1}
\end{figure}

Figure~\ref{e1}a shows  profiles of free surface at various times
obtained from simulations of Dyachenko Eqs. \e{Udef2}, \e{Qdef},
\e{Bintdef2}, \e{Reqn} and \e{Veqn} with  the initial conditions
\e{ini1},\e{inita2a2p01}   and $g=\alpha=0.$ Figures
~\ref{e1}b-\ref{e1}d demonstrate both a persistence in time of poles
originating from  $w=a_1(0), \ w=a_2(0)$     and a formation of
branch cuts at $w=w_\pm.$  Figure~\ref{e1}b shows the positions of
complex singularities of $z_u$ in the complex plane $w\in \C$ at
small times  when the branch cuts originating from $w=w_\pm$  have
small lengths.  Figure  ~\ref{e1}d  shows these positions at larger
times when lengths of these branch cuts increases up to $\sim 1$.
Figure ~\ref{e1}c provides a zoom-in of the left branch cut of
Figure ~\ref{e1}b. The motion of two poles  originating at
$w=a_1(0)$ and $w=a_2(0)$ is shown by thick dots in these Figures.
Branch cuts are numerically approximated in AGH algorithm by a set
of poles according to Eq. \e{approx_cut} with neighboring poles
connected by solid lines in Figure  ~\ref{e1}d. An increase of the
numerical precision results in the increase of number of these
artificial poles approximating the branch cuts. There are several
ways to determine a type of branch point, see e.g. Refs.
\citet{DyachenkoLushnikovKorotkevichJETPLett2014,DyachenkoLushnikovKorotkevichPartIStudApplMath2016}.
Such detailed study of branch point type is however outside the
scope of this paper. We nonly demonstrate a square root branch point
existence
 below in Figure \ref{Residues_r12}a   by a direct fit
of the free surface profile.
 We also note from simulations that at
larger times the poles start absorbing into branch cuts which is
consistent with the assumption of Section
\ref{sec:Newconstantsmotion} that the conservation of the residues
is guaranteed only at small enough times. The study of such
absorbtion is beyond the scope of this paper.

\begin{figure}
\includegraphics[width=0.495\textwidth]{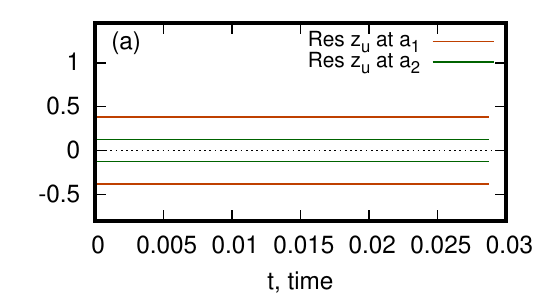}
\includegraphics[width=0.495\textwidth]{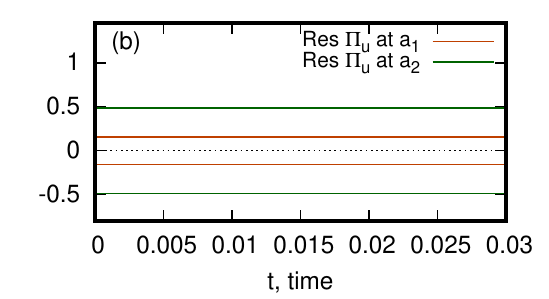}
\includegraphics[width=0.495\textwidth]{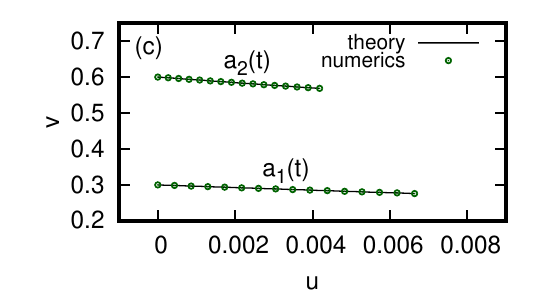}
\includegraphics[width=0.495\textwidth]{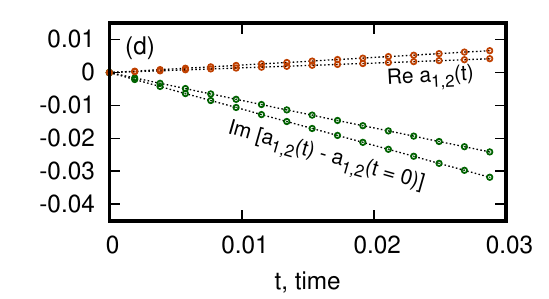}
\includegraphics[width=0.495\textwidth]{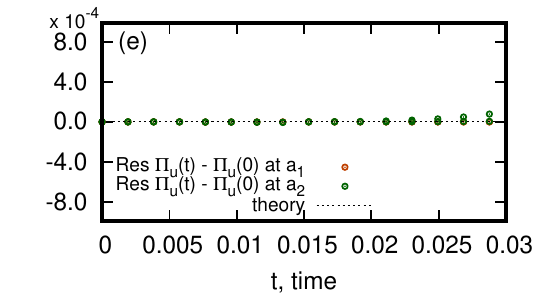}
\includegraphics[width=0.495\textwidth]{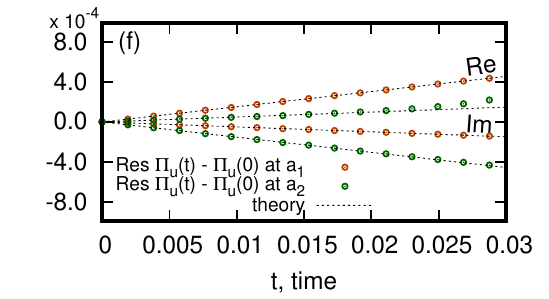}
\caption{ (a) The residues of $z_u$ at $w=a_1(t)$ and $w=a_2(t)$ as
functions of $t$ compared with Eq. \e{Reszw}. (b) The residues of $\Pi_u$ at $w=a_1(t)$ and
$w=a_2(t)$ compared with Eq.   \e{ResPiwt}. (c) Trajectories  of $w=a_1(t)$ and $w=a_2(t)$  in $w$
plane  compared with
the result of the integration of the analytical expression
\e{atzero}. (d) A dependence of real and imaginary parts of $a_1$ and
$a_2$ on $t$ for the same data as in (b). (a)-(d) is shown for the same
simulation  as in Figure \ref{e1} (with $g=0$). (e) and (f) are the same type of plots as  as (a) and (b) except a nonzero gravity
$g=0.04$ is added in the simulation with all other parameters the same as in the simulations of
 Figure \ref{e1}.   } \label{Residues_r11}
\end{figure}

Figure \ref{Residues_r11}a and  \ref{Residues_r11}b demonstrate that
the residues of  both $z_u$ and $\Pi_u$ are the integrals of motions
for $g=0$ fully confirming the analytical results of Eqs. \e{Reszw}
and \e{ResPiwt}. Figure \ref{Residues_r11}c zooms into trajectories
of motion of poles   $w=a_1(t)$ and $w=a_2(t)$ in $w$ plane.  Figure
\ref{Residues_r11}d shows  a time dependence of the pole positions
$w=a_1(t)$ and $w=a_2(t)$ and compares it with the result of the
time integration of Eq. \e{atzero}. The difference between
analytical curves and numerical  ones are nearly  visually
indistinguishable. For that comparison $U$ was calculated
numerically at each moment of time from $R$ and $V$  by using the
definition \e{Udef2} and applying AGH algorithm to recover $U_0(t)$
($U_0$ is defined in Eq. \e{Upoleser2}).
 Only at larger times, when the distance from the branch cuts to either $a_1$ or $a_2$  turns comparable with the spacing between poles approximating branch cut in AGH algorithm, the difference between analytical and numerical values becomes noticeable as expected from the discussion of Section \ref{sec:rationalapproximation}.

Assuming $g=0.04 $ with all other numerical parameters  as above, we obtain
simulation results similar to shown in  Figure \ref{e1} ~because the simulation time remains relatively small so that the
effect of nonzero $g$ is small  for free surface profiles.
However, the residue of $\Pi_u$ is not constant any more but attains the linear dependence on time  as follows
from Eq.   \e{ResPiwt}. Then Figures   \ref{Residues_r11}a and \ref{Residues_r11}b (the case
$g=0$) are replaced by new Figures \ref{Residues_r11}e and \ref{Residues_r11}f (the case
$g=0.04).  $ There is again  the excellent agreement between simulations and the
theoretical curves given by Eqs. \e{Reszw} and \e{ResPiwt}.

\begin{figure}
\includegraphics[width=0.495\textwidth]{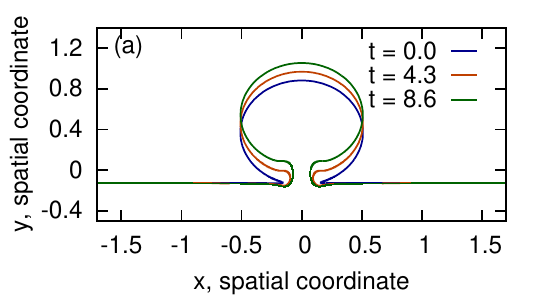}
\includegraphics[width=0.495\textwidth]{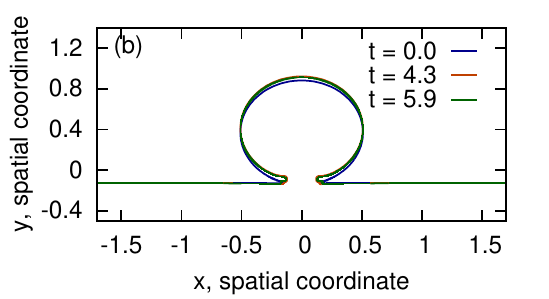}
\includegraphics[width=0.495\textwidth]{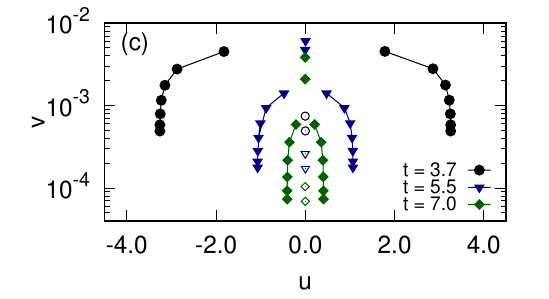}
\includegraphics[width=0.495\textwidth]{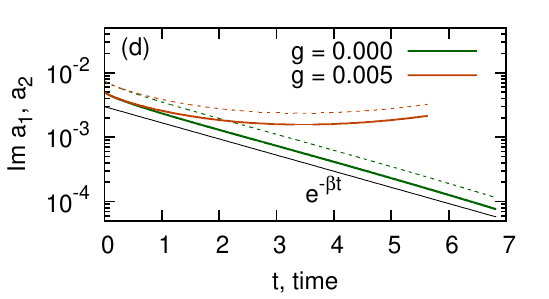}
\includegraphics[width=0.495\textwidth]{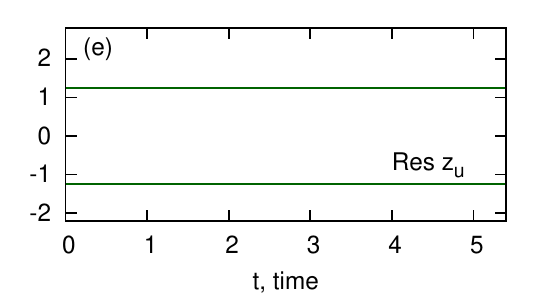}
\includegraphics[width=0.495\textwidth]{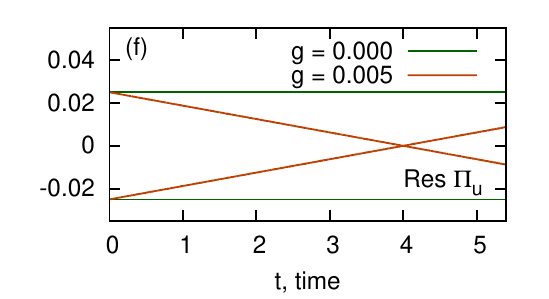}
\caption{Simulations with the initial conditions
\e{ini1},\e{inita2a2p02} and $\alpha=0.$ (a) and (b): profiles of
free surface at different times for $g=0$ and $g=0.005$,
respectively. (c) Positions of two persistent poles (originate at
$w=a_1(0)$ and $w=a_2(0),$  shown by small open circles, triangles
and rhombus) and branch cuts (thick dots and triangles connected by
solid lines) are shown at different $t$ for simulation with $g=0.$
These poles and branch cuts determine the mushroom type shape of
(a). There are other branch cuts well above (not shown) which
determine only a background of the free surface height outside of
the mushroom. (d) Vertical positions for both poles (solid and
dashed lines) for $g=0$ and $g=0.005$ vs. $t$ in log scale. The
exponential dependence $\propto e^{-\beta t}$,
$\beta\simeq0.578329$ is also shown for the comparison. (e) The
residues of $z_u$ at $w=a_1(t)$ and $w=a_2(t)$ extracted from the
simulations are constant in time both for $g=0$ and $g=0.005$ in
agreement with Eq. \e{Reszw}. (f) The residues of $\Pi_u$ at
$w=a_1(t)$ and $w=a_2(t)$ are either constant or liner function of
$t$ depending on $g$ and are visually indistinguishable from   Eq.
\e{ResPiwt}. } \label{r11_surfaceA}
\vspace{-0.1cm}
\end{figure}

We now consider the initial conditions \e{ini1} for another set of numerical values\begin{align}\label{inita2a2p02}
& a_1 (0)= 0.0050\,\I, \, a_2 (0)= 0.0075\, \I, \ q = 1.25\I, \, c = 0.02.
\end{align}
Eqs. \e{wzero1} and \e{inita2a2p02} result in %
$w_+=0.0790677 \ldots + \I \,0.00625 \ldots \quad \text{and} \quad  w_-=-0.0790677 \ldots+ \I\,0.00625\ldots,$
%
i.e. $w_\pm\in\C^+$ in this case as required.
Similar to the previous simulations description of this section, Taylor series expansion of $z_u$ and $\Pi_u$  \e{ini1} at
 $w=w_\pm\in\C^+
$ and $t=0$ reproduces Eqs. \e{Vpole} and \e{Rpole} in the variables $R$ and $V$ with $R_{-1}\ne 0$ and $V_{-1}\ne 0$. Then Theorem 1 of Section \ref{sec:poles} proves that solutions  \e{Vpole} and \e{Rpole} are not persistent in time. At $w=w_\pm$ we again expect a formation of a pair of square root branch points at arbitrary small time $t>0$.  The initial poles at $w=a_1(0)$ and $w=a_2(0)$  are expected to be persistent for at least a finite time duration  according to the results of Section \ref{sec:Newconstantsmotion}.

Figure~\ref{r11_surfaceA}a shows  profiles of free surface at
various times obtained from simulations of Dyachenko Eqs. \e{Udef2},
\e{Qdef}, \e{Bintdef2}, \e{Reqn} and \e{Veqn} with the initial
conditions \e{ini1},\e{inita2a2p02} and $g=\alpha=0.$  Figures
~\ref{r11_surfaceA}b shows a simulation with the same parameters
except   $g=0.005$ and $\alpha=0.$  It is seen at Figures
\ref{r11_surfaceA}a and ~\ref{r11_surfaceA}b that the inial free
surface has a form of disk standing on the nearly flat surface. Then
this disk moves upwards with almost constant velocity (for $g=0$) forming a
mushroom with a narrow neck (stipe). For     $g=0.005$ that upward
motion is quickly suppressed by the nonzero gravity. Figures
~\ref{r11_surfaceA}c demonstrate both a persistence in time of  two
poles originating from  $w=a_1(0), \ w=a_2(0)$     and a
self-similar dynamics of branch cuts originating from $w=w_\pm.$
Figure~\ref{r11_surfaceA}d shows a time dependence of position of two poles
moving strictly in the vertical direction. Contrary to the previous
numerical example, both poles are never absorbed into branch cut and
they are persistent at all times. Figure~\ref{r11_surfaceA}e
and~\ref{r11_surfaceA}f demonstrate that the dynamics of residues of
both $z_u$ and $\Pi_u$ is in full agreement with   Eqs. \e{Reszw}
and \e{ResPiwt} both for $g=0$ and $g\ne 0.$ Log-linear scaling of
Figure~\ref{r11_surfaceA}c also demonstrates    that at large time
an d $g=0$ both  poles and surrounding branch cuts evolve in a
self-similar way (if we rescale with time both $u$ and $v$)
approaching the real line with a spatial scaling $\propto e^{-\beta
t}$, where $\beta\simeq0.578329$ is obtained from the numerical fit
of the curves of Figure ~\ref{r11_surfaceA}d.

\begin{figure}
\includegraphics[width=0.495\textwidth]{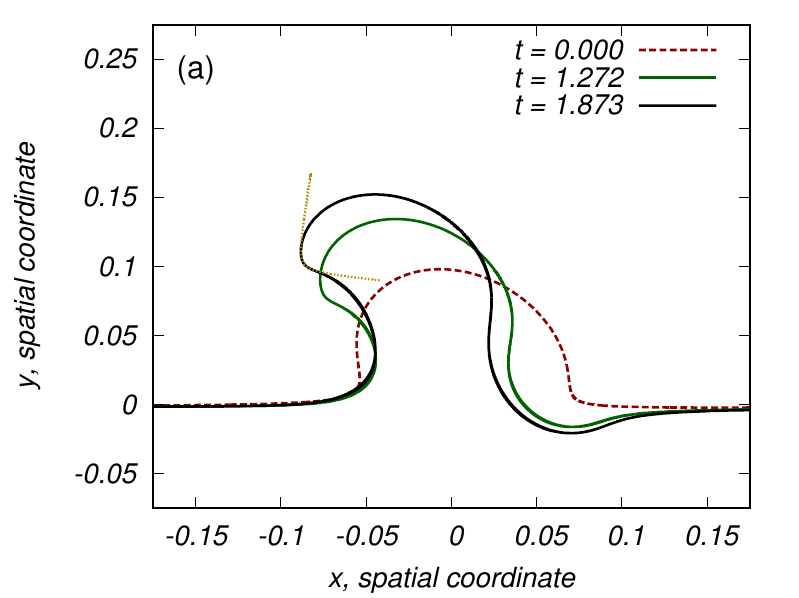}
\includegraphics[width=0.495\textwidth]{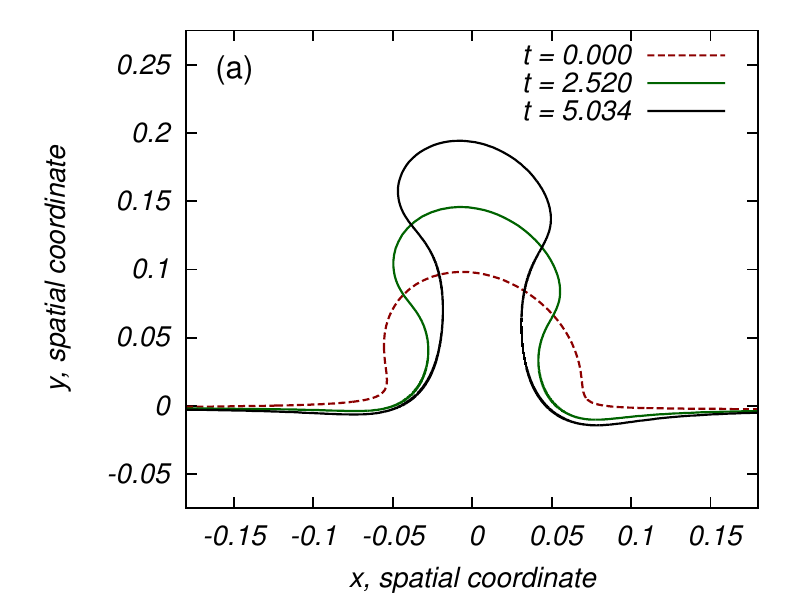}
\includegraphics[width=0.495\textwidth]{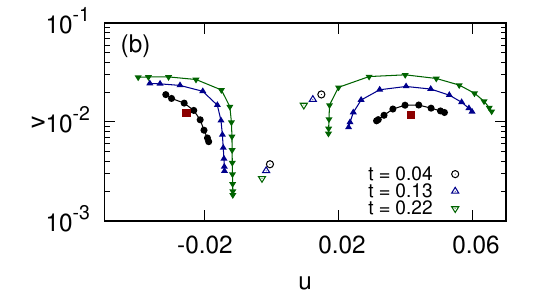}
\includegraphics[width=0.495\textwidth]{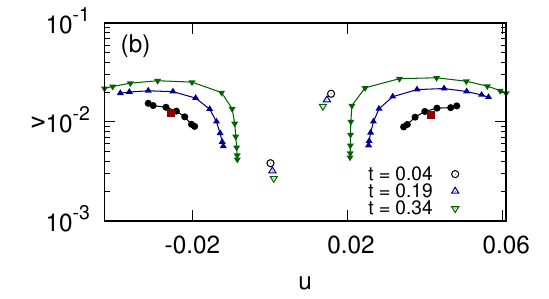}
\includegraphics[width=0.495\textwidth]{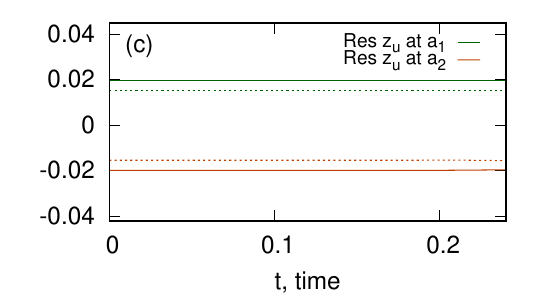}
\includegraphics[width=0.495\textwidth]{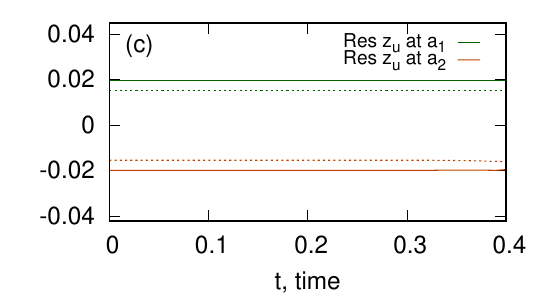}
\caption{ Simulations of the initial conditions
\e{ini1},\e{inita2a2p03},  $g=\alpha=0 $   and either
\e{inita2a2p03a} (left panels) or \e{inita2a2p03b} (right panels).
(a) The shape of the surface. Dotted line on the left panel
shows a fit of  the overturning portion of the wave to the square
root dependence $z=q\,(w-w_c)^{1/2}+z_0$, where fitting parameters are $z_0 = 0.0923+0.0961\, \I$ and $q=0.595329 + 4.48567\, \I$  while $w_c = -0.02348+3.2923  \cdot 10^{-6}\, \I$  is recovered from AGH algorithm
as the position the lowest end of the branch cut. (b) A motion of poles
(small open circles and triangles) and branch cuts (filled circles
and triangles connected by solid lines) in $w$ plane. The small
filled squares show   the point $w=w_\pm$ from Eq. \e{wzero1}. (c)
Residues of $z_u$ extracted from simulations are constant in time in
agreement with Eq. \e{Reszw}. A similar statement is true for
residues of $\Pi_u$ in accordance with Eq. \e{ResPiwt} (not shown).
} \label{Residues_r12}
\end{figure}

Two more sets of the initial conditions \e{ini1} have initial poles
away from the imaginary axis and are given by
\begin{align}\label{inita2a2p03}
& a_1 (0)=0.004\,\I, \, a_2 (0)= 0.016 + 0.020\, \I,  \ q = 0.025e^{\I0.71\pi}
\end{align}
with either \begin{align}\label{inita2a2p03a}
  c=0.03 - 0.02\I
\end{align}
or \begin{align}\label{inita2a2p03b}
c = 0.02.
\end{align}
Eqs. \e{wzero1} and \e{inita2a2p03} result in  %
$w_+=0.0415052  \ldots + \I \,0.0117937 \ldots \quad \text{and} \quad  w_-=-0.0255052 \ldots+ \I\,0.0122063 \ldots,$
%
i.e. $w_\pm\in\C^+$ in these cases as required. Figure
\ref{Residues_r12}a shows  jets  propelled in the direction oblique
to the imaginary axis which is more pronounced in the case
\e{inita2a2p03a} (left panel in Figure  \ref{Residues_r12}a).  The
initial poles at $w=a_1(0)$ and $w=a_2(0)$  are again  persistent in
time with residues obeying Eqs. \e{Reszw}and    \e{ResPiwt} as shown
in Figures \ref{Residues_r12}b and \ref{Residues_r12}c.
Also a fit to the square root dependence shown on left Figure
\ref{Residues_r12}a by a dotted line  corresponds to the square root
branch point at the lowest end of the left branch cut as seen in
Figure \ref{Residues_r12}b.

\subsection{Simulations with second order poles }

\begin{figure}
\includegraphics[width=0.495\textwidth]{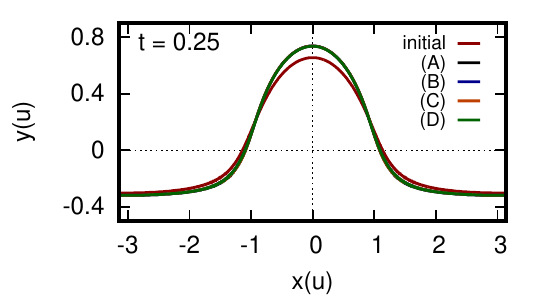}
\includegraphics[width=0.495\textwidth]{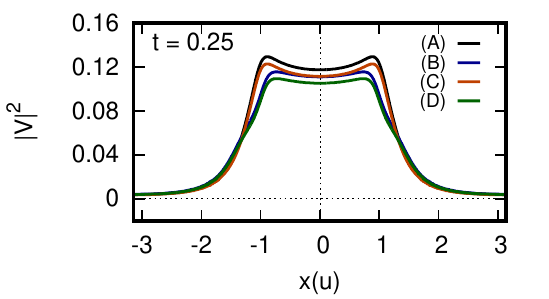}
\caption{ (Left) Free surface profiles resulting from simulations
for the cases (A), (B), (C) and (D) of Eq. \e{galphaset1} at
$t=0.25$ compared with the initial profile at $t=0.$ (Right) $|V|^2$
at the free surface for the same cases. } \label{f1}
\end{figure}

Consider an initial condition in the form of the second order pole at $w=a(0)$ both in  $z_u$ and $\Pi_u$ as follows
\begin{align} \label{zusecond1}
z_u &= 1 + \frac{q}{\cos(w-a(0)) - 1}=1-\frac{q(\zeta^2+1)}{2\cos^2\frac{a}{2}(\zeta-\tan\frac{a}{2})^2}, \\
\Pi_u &=\I c (1 - z_u), \label{Piusecond1}
\end{align}
where $a(0)\in\C^+,$  $c\ne0,q\ne0\in\C$ are the constants and  we used the  identity \e{tansum}.

\begin{figure}
\includegraphics[width=0.495\textwidth]{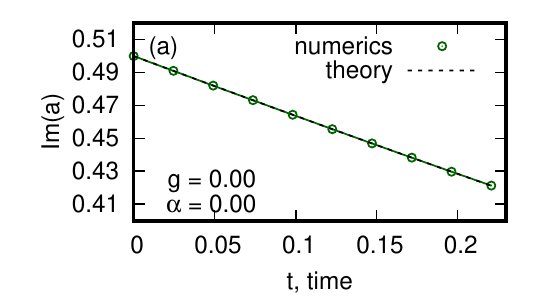}
\includegraphics[width=0.495\textwidth]{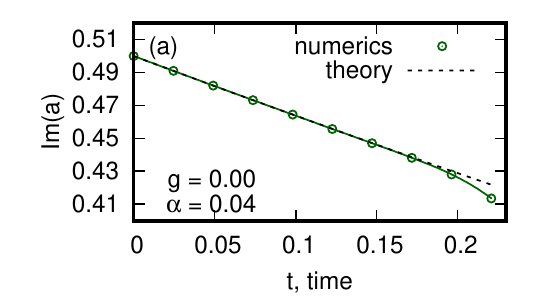}
\includegraphics[width=0.495\textwidth]{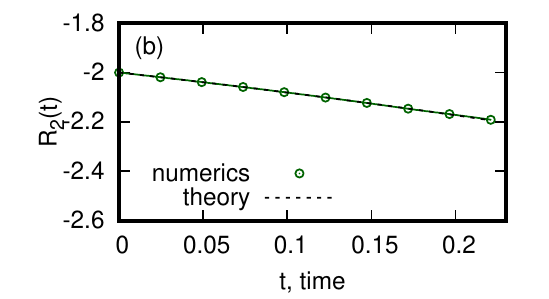}
\includegraphics[width=0.495\textwidth]{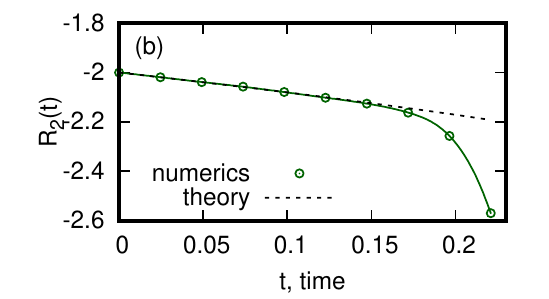}
\includegraphics[width=0.495\textwidth]{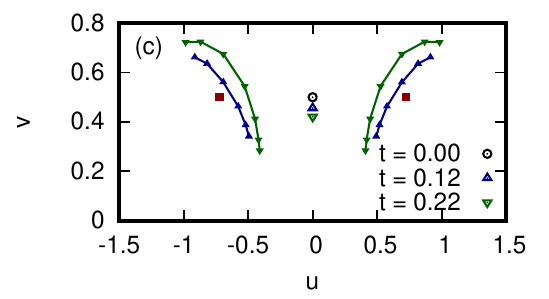}
\includegraphics[width=0.495\textwidth]{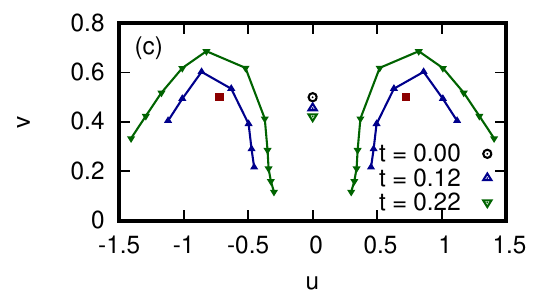}
\caption{ Data extracted from simulations for the cases (A) (left
panels) and (B) (right panels) of Eq. \e{galphaset1}.  (a) A
dependence  of $Im(a)$   on $t$   in Eq. \e{zwPiwn}  compared with
the result of the integration of the analytical expression
\e{atzero3} ($a$ is purely imaginary in these cases). For each
moment of time the location of  $w=a=\I Im(a)$ was found as the
solution of $R(w) = 0$ by the Newton's method. (b) A dependence  of
$R_2$  on $t$   in Eq. \e{zwPiwn} compared with the result of the
integration of the analytical expression \e{R2der3}.  $R_2$  is
purely real  in these cases. In both (a) and (b),    $U_0(t)$ and
$U_1(t)$ were  obtained from AGH algorithm similar to Section
\ref{sec:simplepoles}.
(c) Motion of the pole (small open circle and triangles) and branch cuts (filled triangles connected by solid lines) in $w$ plane for $z(w,t)$.  The small filled squares show   the points $w=w_\pm$ from Eq. \e{wpmsecond01}. 
} \label{fa}
\end{figure}

\begin{figure}
\includegraphics[width=0.495\textwidth]{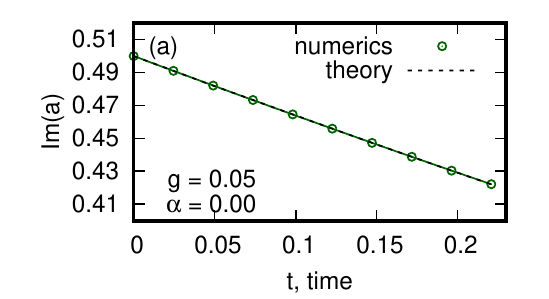}
\includegraphics[width=0.495\textwidth]{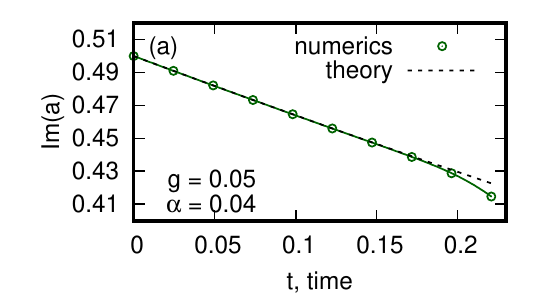}
\includegraphics[width=0.495\textwidth]{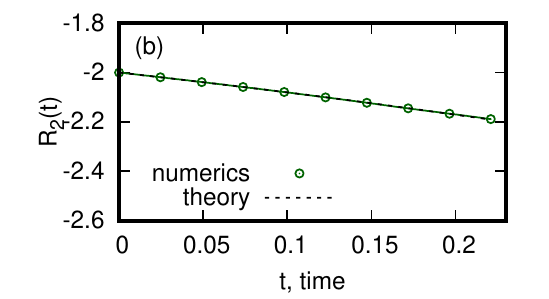}
\includegraphics[width=0.495\textwidth]{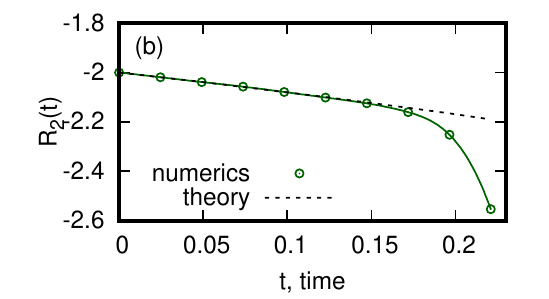}
\caption{ Similar results to Figure \ref{fa}a and \ref{fa}b but for the cases (C) (left panels) and (D) (right panels) of Eq. \e{galphaset1}.
} \label{fb}
\end{figure}

The initial conditions \e{zusecond1} and \e{Piusecond1} together
with Eqs.  \e{RVvar1} and \e{RVvar2} imply    that both $R$ and $V$ are
analytic at $w=a(0)$ for $t=0$  with their Taylor series
coefficients satisfying %
\begin{align} \label{R01V01double}
R_0(0)=R_1(0)=R_3(0)=0, \ R_2(0)=-\frac{1}{2q}\ne0,
V_0(0)=c\ne0, \ V_1(0)=0, \nonumber \\ V_2(0)=\frac{c}{2q} \ \text{at} \ w=a
\end{align}
for $t=0.$ Thus
$R$ has a second order  zero while $V$ is nonzero     at  $w=a(0)$
provided $q\ne 0$ and $c\ne 0$ which corresponds to the case of Eqs.
\e{Rpoleser3} and \e{Vpoleser3}. Then the analytical results of
Section \ref{sec:Newconstantsmotioncapillarity} predict a persistence
of second order poles at $w=a(t)$ of both $z_u$ and $\Pi_u$  for at least a finite duration of time
for arbitrary values of $g$ and $\alpha$. We study four separate
cases $g=\alpha=0;  g= 0, \alpha\ne0; g\ne 0, \alpha=0;g\ne
0, \alpha\ne 0.$
%
%
%


The conformal map \e{zwdef} requires that $z_u\ne 0$ for $w\in\C^-$. Solving for $z_u=0$ in Eq. \e{zusecond1} results in %
\begin{align} \label{wzerosecond1}
w_{\pm}=2\arctan\left [ \frac{2A\pm(1+A^2)\sqrt{q(2-q)}}{2- q[A^2+1]}\right ], \quad A\equiv\tan{\frac{a(0)}{2}}
\end{align}
which provides a restriction on allowed numerical values of $q$ and $a(0)$ to ensure that $w_\pm\in\C^+.$

We choose  numerical values%
\begin{equation} \label{inita2a2psecond01}
   c = 0.5, \quad q = 0.25 \quad  \text{and} \quad  a(0) = 0.5\I
\end{equation}
for all four cases. Eqs. \e{wzerosecond1} and \e{inita2a2psecond01} result in%
\begin{equation} \label{wpmsecond01}
w_+=0.722734 \ldots + \I \,0.5 \quad \text{and} \quad  w_-=-0.722734 \ldots+ \I\,0.5, \end{equation}
i.e. $w_\pm\in\C^+$ in this case as required.
Taylor series expansion of $z_u$ and $\Pi_u$  \e{ini1} at
 $w=w_\pm\in\C^+
$ and $t=0$ reproduces Eqs. \e{Vpole} and \e{Rpole} in the variables
$R$ and $V$ with $R_{-1}\ne 0$ and $V_{-1}\ne 0$. Then Theorem 1 of
Section \ref{sec:poles} proves that solutions  \e{Vpole} and
\e{Rpole} are not persistent in time. Similar to the discussion of
Section \ref{sec:simplepoles}, we expect a formation of a  pair of
square root branch points at an arbitrary small time $t>0$.

Figures~\ref{f1}a and ~\ref{f1}b  show  profiles of free surface and $|V|^2$ obtained from simulations of Dyachenko Eqs. \e{Udef2}, \e{Qdef}, \e{Bintdef2}, \e{Reqn} and \e{Veqn} with the  initial conditions \e{zusecond1},\e{Piusecond1},\e{inita2a2psecond01} and four particular cases   %
\begin{equation} \label{galphaset1}
(A)  g=\alpha=0;\ (B) g= 0,\,\alpha=0.04; \ (C) g=0.05, \alpha=0; \   \text{and}\  (D) g=0.05, \alpha=0.04.
\end{equation}
Figures ~\ref{fa}a and \ref{fa}b show time dependencies of the
second order pole of both    $z_w(w,t) $ and $\Pi_w(w,t)$  at
$w=a(t)$ as well as  the coefficient $R_2(t)$ of Taylor series
\e{Rpoleser3} (also enter into Eqs. \e{zwPiwn}) compared with a time
integration of Eqs.  \e{atzero3} and \e{R2der3}. It confirms a
persistence in time of second order poles originating from  $w=a(0)
$ for the initial conditions \e{zusecond1} and \e{Piusecond1}. We
also recovered  $V_1(t)$  from simulations (not shown in Figures)
which together with $R_2(t)$ allowed to confirm the integral of
motion \e{R2V1const}.

Figure~\ref{fa}c shows the positions of complex singularities of $z$
in the complex plane $w\in \C.$   The branch cuts form at arbitrary
small time $t>0$ from the points  $w=w_\pm$ \e{wpmsecond01}. It is
seen that the nonzero surface tension on the right panel of
Figure~\ref{fa}c results in a significantly faster extension of
these branch cuts   compared with the zero surface tension case on
the left panel. This is consistent with the results of Ref. \cite{DyachenkoNewell2016} that an addition of surface tension results in quick approach of singularities to the real line.    Similar to simulations of Section
\ref{sec:simplepoles},  at larger times the poles start absorbing
into branch cuts. The deviation between analytical and numerical
results in right panels of   Figures~\ref{fa}a and \ref{fa}b at
later times is due to the faster approach of branch cuts to the pole
position for $\alpha\ne 0$ thus resulting in AGH algorithm to loose
the numerical precision  as expected from the discussion of Section
\ref{sec:rationalapproximation}. We also note that our use of $z$
(instead of using $z_u$ in Section \ref{sec:simplepoles}) to obtain
Figures~\ref{fa}a-\ref{fa}c is due to the convenience of recovering
simple poles in AGH algorithm compared with the second order poles.
Indeed, $z$ has the first order  pole at $w=a$ as obtained from the
integration of Eq. \e{zwPiwn} over $w$. Generally such integration
produces also a logarithmic branch point  $w=a$ from the simple pole
in Eq. \e{zwPiwn} which would imply a formation of multiple poles
approximating that branch point by AGH algorithm in
Figure~\ref{fa}c. However, the particular initial conditions
\e{zusecond1} and \e{Piusecond1} imply through Eqs.
\e{R2V1const},\e{Reszwm}-\e{ReszwPiw2},\e{R01V01double}  that
$V_1(t)=R_3(t)=0,$ i.e.
$\underset{w=a}{Res}(\Pi_w)=\underset{w=a}{Res}(z_w)=0$ thus
removing a logarithmic branch point   $w=a.$  The absence a
logarithmic branch point   $w=a$    in Figure \ref{fa}c   also
provides another confirmation of the persistence of the second order
pole in $z_u$ at    $w=a$ and the validity of the motion integrals
\e{R2V1const} and \e{Reszwm}-\e{ReszwPiw2}.

Figure \ref{fb} shows results similar Figures ~\ref{fa}a and \ref{fa}b  but with $g\ne 0. $  The positions of complex singularities are not shown because they are nearly the same as in Figure ~\ref{fa}c.

We  conclude that in this section we verified with a high numerical precision a conservation of both complex integrals of motion  of Section  \ref{sec:Newconstantsmotion} and  all three independent complex integrals of motions for the second order pole case of Section \ref{sec:Newconstantsmotioncapillarity}.

\section{Conclusion and Discussion}
\label{sec:conclusion}

The main result of this paper is the existence of   new integrals of motion in free surface hydrodynamics. These integrals are closely tied to the existence of solutions with poles of the first and the second orders in both
$z_w$ and $\Pi_w.$ The residues of $z_w$ are the integral of motion while residues of $\Pi_w$ are the linear function of time for nonzero gravity turning into the integrals of motion for zero gravity. The residues of $z_w$  at different points commute with each other in
the sense of underlying non-canonical Hamiltonian dynamics.  It provides an argument in support of the conjecture of complete integrability
of free surface hydrodynamics in deep water. We also suggested to treat the analytical continuation of the free surface dynamics outside of the physical fluid as the phantom hydrodynamics on the multi-sheet Riemann surface. That phantom hydrodynamics allows a generalized Kelvin theorem. We expect that generally a number of sheets will be infinite with generic solutions to involve  poles and square root branch points in multiple sheets.

For future work we suggest an extension to the general case of the poles of arbitrary order in
$z_w$ and $\Pi_w$ to count the total number of independent integrals
of motion. We propose to also study the expected pole solutions in other (nonphysical) sheets of Riemann surface. The commutativity properties between different integral of motion need to be studied in the general case.

\section{Acknowledgements.}

 The  work of A.I.D., P.M.L.   and V.E.Z.  was   supported by the state assignment "``Dynamics of the complex systems"".
 The  work of
P.M.L.  was   supported by the National Science Foundation, grant
DMS-1814619. The work of S.A.D. was supported by the National
Science Foundation, grant number DMS-1716822. The work of V.E.Z. was
supported by the National Science Foundation, grant number
DMS-1715323. The work of A.I.D. and V.E.Z. described in sections 2,
3 and 6 was supported by the Russian Science Foundation, grant
number 19-72-30028.


\end{document}